\documentstyle[aps,epsfig]{revtex}

\begin{document}
\preprint{TRI-PP-96-19}
\draft

\title{Validity of Certain Soft Photon Amplitudes}

\author{Mark Welsh}
\address{Department of Physics, University of Victoria, Victoria, B.C., 
Canada V8W 3P6}
\author{Harold W. Fearing}
\address{TRIUMF, 4004 Wesbrook Mall, Vancouver, B.C., Canada V6T 2A3}

\date{TRI-PP-96-19,\hspace{3mm} June 18, 1996}

\maketitle

\begin{abstract}
Certain soft photon amplitudes which have been recently suggested as 
alternatives to the usual Low form of the soft photon approximation 
are studied and it
is demonstrated that problems exist in their relation to the corresponding 
non-radiative amplitude. The non-radiative amplitude, which is an input 
to soft photon calculations, is in certain cases required to be evaluated 
outside of its physical phase space region. 
Also, for the case of two-body identical particle bremsstrahlung 
processes, the symmetrized or antisymmetrized form of these soft photon 
amplitudes cannot be written in terms of the symmetrized or antisymmetrized 
amplitude for the non-radiative process. It is found that the usual Low form 
of the soft photon theorem is essentially unaffected by these problems.
\end{abstract}



\section{Introduction}
\label{sectI}

Bremsstrahlung processes, particularly proton-proton bremsstrahlung, 
have long been studied as a method of assessing the
importance of off-shell effects in low and intermediate energy hadronic 
scattering. There have been two main theoretical approaches:
non-relativistic potential models 
\cite{fearing86,brown91,herrmann92,katsog93,jetter94,jetter95,eden96} 
which include off-shell
effects explicitly, and the soft photon approximation 
\cite{low58,nyman67,fearing72b,fearing73,fearing87,liou95} 
which is written
in terms of only on-shell information about the non-radiative
scattering process. Soft photon amplitudes therefore give information
about off-shell effects only through any discrepancy between their
prediction for the bremsstrahlung spectrum and experimental
measurements, and even then there is an ambiguity in that some of the 
discrepancy could arise from higher order on-shell effects. 

For the proton-proton bremsstrahlung process it had been found in the
past that the Low soft photon approximation gave a good
description of the older data\cite{rogers80}, suggesting that 
off-shell effects are
small. The more recent 280~MeV TRIUMF experiment \cite{michaelian90} provided
measurements not only of photon spectra but also of polarization
observables. This data showed some disagreement with the soft photon
prediction \cite{fearing87}, indicating for the first time the presence
of non-trivial off-shell behavior in the $pp$~elastic scattering
process. 

Although most soft photon applications have used the Low \cite{low58} 
approach, it is well known that the derivation of the soft photon 
approximation is not
unique. Different choices lead to soft photon amplitudes which differ at 
${\cal O}(k)$. Recently Liou, Timmermans and Gibson \cite{liou95} 
have suggested an
alternate form for the soft photon amplitude. The claim is made that
this ``Two-u-Two-t-special'' ({\sc TuTts}) amplitude provides better
agreement with $pp\gamma$ data at all energies than the traditional
Low amplitude. This success is contrasted in that paper with the
dramatic failure to describe the data of another alternative soft photon 
amplitude, the ``Two-s-Two-t-special'' ({\sc TsTts}) amplitude. 

In this paper we shall investigate two problems which can arise in the
application of soft photon amplitudes to particular processes. The
first of these, which we shall call the phase space problem, concerns
the expression of a soft photon amplitude solely in terms of
measurable information about the on-shell non-radiative process. The
second, the antisymmetrization problem, concerns the inability to
write the correctly antisymmetrized $pp\gamma$ soft photon amplitude
in terms of the measured, antisymmetric $pp$~elastic amplitude. The
usual Low form for the radiative amplitude will be shown to be immune
to these difficulties, while the {\sc TuTts} and {\sc TsTts} amplitudes fall
victim to one or both of the problems. 

These problems both have analogs in the case of spinless two-body
bremsstrahlung processes. We begin by studying the problems in that
algebraically simpler context, deriving the spinless forms of the Low,
{\sc TuTts} and {\sc TsTts} amplitudes in section~\ref{sectII}. In
section~\ref{sectIII} we consider the phase space problem while in
section~\ref{sectIV} we treat the symmetrization of identical particle
spinless bremsstrahlung processes. In section~\ref{sectV} we extend
our results to proton-proton bremsstrahlung where the elastic and
radiative amplitudes must be written in antisymmetric form, and
present some illustrative examples of the problems discussed.

\section{Spin-0 Amplitudes}
\label{sectII}

We begin with two-body spin zero scattering and first review the derivation 
of the Low 
\cite{low58} form of the soft photon approximation as well as the 
``Two-s-Two-t-special'' ({\sc TsTts}) and ``Two-u-Two-t-special'' 
({\sc TuTts}) forms suggested by Liou and collaborators \cite{liou95}. 
The problems in which we are interested
may be considered within this algebraically simpler spinless framework,
and then carried over with little modification to the more physically
interesting case of nucleon-nucleon bremsstrahlung. 

We define $A(s,t;p_1^2,p_2^2,p_3^2,p_4^2)$ to be the amplitude
describing the non-radiative scattering of particles of mass $m_1$ 
and $m_2$ into
a final state composed of masses $m_3$ and $m_4$ with $s$ and
$t$ the usual Mandelstam variables. The $p_i$ are the four-momenta of the 
various particles and the variables $s$ and $t$, 
and others to be defined later, 
are considered to be functions of these four-momenta. Contact with the 
physical non-radiative amplitude, which can be evaluated from measured phase 
shifts, is made by going to the on-shell limit, $p_i^2=m_i^2$.
The Low form of the soft
photon amplitude may then be constructed as follows. One writes the
contribution of radiation from the external charged particles to the
radiative amplitude in terms of off-shell evaluations of the
non-radiative scattering amplitude $A()$. 
\begin{eqnarray}
{\cal M}_{\mbox{\tiny ext}}^\mu\epsilon_\mu & = & 
eQ_3\frac{p_3\cdot\epsilon}{p_3\cdot k} 
A\left( \overline{s}+k\cdot(p_3+p_4),\overline{t}-k\cdot(p_1-p_3); 
m_1^2, m_2^2, m_3^2+2k\cdot p_3, m_4^2\right) \nonumber \\ & & + 
eQ_4\frac{p_4\cdot\epsilon}{p_4\cdot k} 
A\left( \overline{s}+k\cdot(p_3+p_4),\overline{t}-k\cdot(p_2-p_4); 
m_1^2, m_2^2, m_3^2, m_4^2+2k\cdot p_4\right) \nonumber \\ & & - 
eQ_1\frac{p_1\cdot\epsilon}{p_1\cdot k} 
A\left( \overline{s}-k\cdot(p_1+p_2),\overline{t}-k\cdot(p_1-p_3); 
m_1^2-2k\cdot p_1, m_2^2, m_3^2, m_4^2\right) \nonumber \\ & & - 
eQ_2\frac{p_2\cdot\epsilon}{p_2\cdot k} 
A\left( \overline{s}-k\cdot(p_1+p_2),\overline{t}-k\cdot(p_2-p_4); 
m_1^2, m_2^2-2k\cdot p_2, m_3^2, m_4^2\right).
\label{eq:lowextern}
\end{eqnarray}
Here $Q_i$ are the 
charges of the various particles and $k^\mu$ and $\epsilon^\mu$ are the
photon momentum and polarization vector. The non-radiative amplitude
$A()$ is written with each of the charged legs in turn taken off-shell
due to the emission of the photon, i.e., we use the same functional form as 
the on-shell, non-radiative amplitude considered as a function of 
the $p_i$, but 
evaluate the $p_i$ at the radiative point satisfying $p_1+p_2=p_3+p_4+k$. We 
have chosen to express this off-shell 
behavior in terms of the average Mandlestam variables
$\overline{s}\equiv\mbox{\small${1\over 2}$}(p_1+p_2)^2+
\mbox{\small${1\over 2}$}(p_3+p_4)^2$ and 
$\overline{t}\equiv\mbox{\small${1\over 2}$}(p_1-p_3)^2+
\mbox{\small${1\over 2}$}(p_2-p_4)^2$. Other choices
can be made for the variables and such choices are at this stage  
entirely equivalent but would later give rise to soft
photon amplitudes differing by terms of ${\cal O}(k)$. 

Following Low \cite{low58}, the occurrences of the non-radiative
amplitude $A()$ in this radiative amplitude are expanded in
powers of $k^\mu$ about the point with explicit dependencies on
$k^\mu$ set to zero---in our current example we expand about
$A(\overline{s},\overline{t};m_1^2,m_2^2,m_3^2,m_4^2)$. Only the
leading two powers in this expansion are retained since it has been
shown \cite{fearing73,bellvanroyen69,ferrari68} that the soft photon
approximation is ambiguous in its prediction of higher orders in the
power of $k^\mu$ expansion due to the ambiguity in choice of
expansion point. The truncated expansion of Eq.~(\ref{eq:lowextern}) is
\begin{eqnarray}
{\cal M}_{\mbox{\tiny ext}}^\mu\epsilon_\mu & = & 
\left[eQ_3\frac{p_3\cdot\epsilon}{p_3\cdot k} 
\left(1+k\cdot(p_3+p_4)\frac{\partial}{\partial\overline{s}} 
-k\cdot(p_1-p_3)\frac{\partial}{\partial\overline{t}} 
+2k\cdot p_3\frac{\partial}{\partial m_3^2}\right)\right.\nonumber\\ 
&&\quad+eQ_4\frac{p_4\cdot\epsilon}{p_4\cdot k} 
\left(1+k\cdot(p_3+p_4)\frac{\partial}{\partial\overline{s}} 
-k\cdot(p_2-p_4)\frac{\partial}{\partial\overline{t}} 
+2k\cdot p_4\frac{\partial}{\partial m_4^2}\right)\nonumber\\ 
&&\quad-eQ_1\frac{p_1\cdot\epsilon}{p_1\cdot k} 
\left(1-k\cdot(p_1+p_2)\frac{\partial}{\partial\overline{s}} 
-k\cdot(p_1-p_3)\frac{\partial}{\partial\overline{t}} 
-2k\cdot p_1\frac{\partial}{\partial m_1^2}\right)\nonumber\\ 
&&\quad\left.-eQ_2\frac{p_2\cdot\epsilon}{p_2\cdot k} 
\left(1-k\cdot(p_1+p_2)\frac{\partial}{\partial\overline{s}} 
-k\cdot(p_2-p_4)\frac{\partial}{\partial\overline{t}} 
-2k\cdot p_2\frac{\partial}{\partial m_2^2}\right)\right]\nonumber\\ 
&&\qquad\qquad\qquad\qquad\qquad\qquad\qquad\qquad
\times A\left(\overline{s},\overline{t};m_1^2,m_2^2,m_3^2,m_4^2\right).
\end{eqnarray}
This truncated form is no longer gauge invariant. Gauge invariance may
be reimposed by the addition of a term 
${\cal M}_{\mbox{\tiny int}}^\mu\epsilon_\mu$ which is independent of $k$ 
and is presumed
to have its physical origin in photon emission from internal charged
lines in the scattering process. The gauge invariance constraint is 
$({\cal M}_{\mbox{\tiny ext}}^\mu+{\cal M}_{\mbox{\tiny int}}^\mu)k_\mu
\equiv0$, which in this case implies an internal contribution of the
form 
\begin{eqnarray}
{\cal M}_{\mbox{\tiny int}}^\mu\epsilon_\mu & = & 
-\Bigl[e(Q_1+Q_2+Q_3+Q_4)(p_3+p_4)\cdot\epsilon
\frac{\partial}{\partial\overline{s}}
+e(Q_1-Q_2-Q_3+Q_4)(p_1-p_3)\cdot\epsilon
\frac{\partial}{\partial\overline{t}}\nonumber\\
&&\quad+2e\Bigl(Q_1p_1\cdot\epsilon\frac{\partial}{\partial m_1^2}
+Q_2p_3\cdot\epsilon\frac{\partial}{\partial m_2^2}
+Q_2p_3\cdot\epsilon\frac{\partial}{\partial m_3^2}
+Q_2p_3\cdot\epsilon\frac{\partial}{\partial m_4^2}\Bigr)\Bigr]
\nonumber\\&&\qquad\qquad\qquad\qquad\qquad\qquad\qquad\qquad
\times A\left(\overline{s},\overline{t};m_1^2,m_2^2,m_3^2,m_4^2\right).
\label{eq:lowintern}
\end{eqnarray}
There is an ambiguity here in the choice of internal radiation
contribution since any independently gauge invariant term could also be
added to the radiative amplitude at this point. 

The soft photon amplitude is the sum of the external and internal
contributions.
\begin{eqnarray}
{\cal M}^\mu_{\mbox{\tiny soft}}\epsilon_\mu&\equiv&
({\cal M}^\mu_{\mbox{\tiny ext}}+{\cal M}^\mu_{\mbox{\tiny int}})
\epsilon_\mu\nonumber\\
&=&\left\{ eQ_3\frac{p_3^\mu}{p_3\cdot k} + eQ_4\frac{p_4^\mu}{p_4\cdot k} 
-eQ_1\frac{p_1^\mu}{p_1\cdot k}-eQ_2\frac{p_2^\mu}{p_2\cdot k}\right. 
\nonumber \\
& & + eQ_3 \left[ \frac{p_4\cdot k}{p_3\cdot k}p_3^\mu-p_4^\mu \right] 
\frac{\partial}{\partial \overline{s}}
- eQ_3 \left[ \frac{p_1\cdot k}{p_3\cdot k}p_3^\mu-p_1^\mu \right] 
\frac{\partial}{\partial \overline{t}} \nonumber \\
& & + eQ_4 \left[ \frac{p_3\cdot k}{p_4\cdot k}p_4^\mu-p_3^\mu \right] 
\frac{\partial}{\partial \overline{s}}
- eQ_4 \left[ \frac{p_2\cdot k}{p_4\cdot k}p_4^\mu-p_2^\mu \right] 
\frac{\partial}{\partial \overline{t}} \nonumber \\
& & + eQ_1 \left[ \frac{p_2\cdot k}{p_1\cdot k} p_1^\mu-p_2^\mu \right] 
\frac{\partial}{\partial \overline{s}}
- eQ_1 \left[ \frac{p_3\cdot k}{p_1\cdot k} p_1^\mu-p_3^\mu \right] 
\frac{\partial}{\partial \overline{t}} \nonumber \\
& & \left. 
+ eQ_2 \left[ \frac{p_1\cdot k}{p_2\cdot k}p_2^\mu-p_1^\mu \right] 
\frac{\partial}{\partial \overline{s}}
- eQ_2 \left[ \frac{p_4\cdot k}{p_2\cdot k}p_2^\mu-p_4^\mu \right] 
\frac{\partial}{\partial \overline{t}} \right\}\epsilon_\mu \nonumber \\
& & \hspace{30mm} \times A(\overline{s}, \overline{t}, m_1^2, m_2^2, 
m_3^2, m_4^2)
\label{eq:lowsoft}
\end{eqnarray}
This amplitude is usually referred to as the Low choice, although it
differs slightly from the construction used in Low's original paper
\cite{low58}. It is distinguished by the choice of a single expansion
point for the non-radiative amplitude.

Recently Liou, Timmermans and Gibson \cite{liou95} have considered 
choices of expansion point which limit the explicit $k^\mu$ dependence
of the non-radiative amplitude $A()$ in Eq.~(\ref{eq:lowextern}) to
its invariant mass arguments. This removes the derivatives with
respect to $s$-type and $t$-type variables from the resulting soft
photon amplitude. The authors of \cite{liou95} suggest that this
property makes the soft photon approximation more suitable for
application to processes thought to be dominated by $s$ or $t$ 
channel resonances. To obtain their result, one is required to use a different
pair of radiative variables for expansion of the external radiation 
contribution of each charged particle. Eq.~(\ref{eq:lowextern}) now
takes the form
\begin{eqnarray}
{\cal M}^\mu\epsilon_\mu & = & 
eQ_3\frac{p_3\cdot\epsilon}{p_3\cdot k} 
A\left(s_{12},t_{24}; 
m_1^2, m_2^2, m_3^2+2k\cdot p_3, m_4^2\right) \nonumber \\ & & + 
eQ_4\frac{p_4\cdot\epsilon}{p_4\cdot k} 
A\left(s_{12},t_{13};
m_1^2, m_2^2, m_3^2, m_4^2+2k\cdot p_4\right) \nonumber \\ & & - 
eQ_1\frac{p_1\cdot\epsilon}{p_1\cdot k} 
A\left(s_{34},t_{24};
m_1^2-2k\cdot p_1, m_2^2, m_3^2, m_4^2\right) \nonumber \\ & & - 
eQ_2\frac{p_2\cdot\epsilon}{p_2\cdot k} 
A\left(s_{34},t_{13};
m_1^2, m_2^2-2k\cdot p_2, m_3^2, m_4^2\right)
\label{eq:tsttsextern}
\end{eqnarray}
where we have defined $s_{12}\equiv(p_1+p_2)^2$,
$s_{34}\equiv(p_3+p_4)^2$, $t_{13}\equiv(p_1-p_3)^2$,
$t_{24}\equiv(p_2-p_4)^2$ and where again $A()$ is considered an implicit
function of the four-momenta $p_i$. By expanding Eq.~(\ref{eq:tsttsextern}) 
about the point $k^\mu = 0$, i.e. expanding in the explicit $k^\mu$ 
dependence, and
truncating after the leading two terms  
in $k^\mu$, and reimposing gauge invariance we have the result
\begin{eqnarray}
{\cal M}_{\mbox{\tiny soft}}'\cdot\epsilon & = & 
eQ_3\frac{p_3\cdot\epsilon}{p_3\cdot k} 
A\left(s_{12},t_{24}; 
m_1^2, m_2^2, m_3^2, m_4^2\right) + 
eQ_4\frac{p_4\cdot\epsilon}{p_4\cdot k} 
A\left(s_{12},t_{13};
m_1^2, m_2^2, m_3^2, m_4^2\right) \nonumber \\ & & - 
eQ_1\frac{p_1\cdot\epsilon}{p_1\cdot k} 
A\left(s_{34},t_{24};
m_1^2, m_2^2, m_3^2, m_4^2\right) - 
eQ_2\frac{p_2\cdot\epsilon}{p_2\cdot k} 
A\left(s_{34},t_{13};
m_1^2, m_2^2, m_3^2, m_4^2\right) \nonumber \\ & & -B^\mu\epsilon_\mu
\label{eq:tsttspresoft}
\end{eqnarray}
where, due to gauge invariance, $B^\mu$ must satisfy the constraint
\begin{eqnarray}
B^\mu k_\mu&=&
eQ_3A\left(s_{12},t_{24};m_1^2,m_2^2,m_3^2,m_4^2\right)
+eQ_4A\left(s_{12},t_{13};m_1^2,m_2^2,m_3^2,m_4^2\right)\nonumber\\&&
-eQ_1A\left(s_{34},t_{24};m_1^2,m_2^2,m_3^2,m_4^2\right)
-eQ_2A\left(s_{34},t_{13};m_1^2,m_2^2,m_3^2,m_4^2\right).
\end{eqnarray}
In order to obtain the ``Two-s-Two-t-special'' ({\sc TsTts})
amplitude of \cite{liou95} we must choose $B^\mu$ itself to have the form 
\begin{eqnarray}
B^\mu&\equiv&\frac{(p_1+p_2)^\mu}{(p_1+p_2)\cdot k}
\Bigl[eQ_3A\left(s_{12},t_{24};m_1^2,m_2^2,m_3^2,m_4^2\right)
+eQ_4A\left(s_{12},t_{13};m_1^2,m_2^2,m_3^2,m_4^2\right)\nonumber\\&&
\qquad\qquad\qquad-eQ_1A\left(s_{34},t_{24};m_1^2,m_2^2,m_3^2,m_4^2\right)
-eQ_2A\left(s_{34},t_{13};m_1^2,m_2^2,m_3^2,m_4^2\right)\Bigr].
\label{eq:tsttsB}
\end{eqnarray}
The {\sc TsTts} amplitude is then 
\begin{eqnarray}
{\cal M}_{\mbox{\small\sc TsTts}}\cdot\epsilon & = & 
eQ_3\left(\frac{p_3\cdot\epsilon}{p_3\cdot k} 
-\frac{(p_3+p_4)\cdot\epsilon}{(p_3+p_4)\cdot k}\right)
A\left(s_{12},t_{24}; 
m_1^2, m_2^2, m_3^2, m_4^2\right) \nonumber \\ & & + 
eQ_4\left(\frac{p_4\cdot\epsilon}{p_4\cdot k} 
-\frac{(p_3+p_4)\cdot\epsilon}{(p_3+p_4)\cdot k}\right)
A\left(s_{12},t_{13}; 
m_1^2, m_2^2, m_3^2, m_4^2\right) \nonumber \\ & & -
eQ_1\left(\frac{p_1\cdot\epsilon}{p_1\cdot k} 
-\frac{(p_1+p_2)\cdot\epsilon}{(p_1+p_2)\cdot k}\right)
A\left(s_{34},t_{24}; 
m_1^2, m_2^2, m_3^2, m_4^2\right) \nonumber \\ & & -
eQ_2\left(\frac{p_2\cdot\epsilon}{p_2\cdot k} 
-\frac{(p_1+p_2)\cdot\epsilon}{(p_1+p_2)\cdot k}\right)
A\left(s_{34},t_{13}; 
m_1^2, m_2^2, m_3^2, m_4^2\right)
\label{eq:tsttssoft}
\end{eqnarray}
where we have employed the relation
\[\frac{(p_3+p_4)\cdot\epsilon}{(p_3+p_4)\cdot k}=
\frac{(p_1+p_2)\cdot\epsilon}{(p_1+p_2)\cdot k}.\]

The form of $B^\mu$ in Eq.~(\ref{eq:tsttsB}) is troubling since it
appears to have a $1/k^\mu$ dependence and yet is assumed to represent
terms which would arise in a perturbative treatment due to radiation 
from internal charged lines. 
Such internal radiation is known from perturbation theory
arguments to give contributions regular in $k^\mu$ as
$k^\mu\rightarrow0$ \cite{adler66}. By expansion of the occurrences of the
non-radiative amplitude $A()$ in Eq.~(\ref{eq:tsttsB}) about a common
point, say $A(s_{12},t_{13};m_1^2,m_2^2,m_3^2,m_4^2)$, one can show
that so long as the charge condition $Q_1=Q_3$, $Q_2=Q_4$ is satisfied the
apparent $1/k^\mu$ dependence vanishes. This charge condition holds 
for the elastic scattering processes in which we are interested. 
For processes where the charge condition is not satisfied the
{\sc TsTts} amplitude would contain unphysical terms in its internal
radiation part, and so would be ill-defined. 

The remaining soft photon amplitude which we shall later use as an
example is the ``Two-u-Two-t-special'' ({\sc TuTts}) amplitude of
Ref.~\cite{liou95}. In constructing it we first note that the
non-radiative amplitude may be parameterized in terms of the
Mandelstam variables $u$ and $t$, rather than $s$ and $t$. We define a
function $A'(u,t)\equiv A(s,t)$ subject to the
constraint $s+t+u=\sum_{i=1}^4m_i^2$. Thus $A'(u,t)$ is just $A(s,t)$ with $s$
replaced by $\sum_{i=1}^4m_i^2-u-t$. For the on-shell elastic process,
$A'(u,t)$ is of course identical to $A(s,t)$. However it is a different
function of the $p_i$ and so has a different value than $A(s,t)$ when they are
evaluated using the radiative $p_i$ instead of the non-radiative ones. 

The off-shell external radiation
amplitude of Eqs.~(\ref{eq:lowextern}) and~(\ref{eq:tsttsextern})
which forms the starting point of the soft photon approximation may be
written in terms of this function as
\begin{eqnarray}
{\cal M}^\mu\epsilon_\mu & = & 
eQ_3\frac{p_3\cdot\epsilon}{p_3\cdot k} 
A'\left(u_{14},t_{24}; 
m_1^2, m_2^2, m_3^2+2k\cdot p_3, m_4^2\right) \nonumber \\ & & + 
eQ_4\frac{p_4\cdot\epsilon}{p_4\cdot k} 
A'\left(u_{23},t_{13};
m_1^2, m_2^2, m_3^2, m_4^2+2k\cdot p_4\right) \nonumber \\ & & - 
eQ_1\frac{p_1\cdot\epsilon}{p_1\cdot k} 
A'\left(u_{23},t_{24};
m_1^2-2k\cdot p_1, m_2^2, m_3^2, m_4^2\right) \nonumber \\ & & - 
eQ_2\frac{p_2\cdot\epsilon}{p_2\cdot k} 
A'\left(u_{14},t_{13};
m_1^2, m_2^2-2k\cdot p_2, m_3^2, m_4^2\right)
\end{eqnarray}
where, as in the {\sc TsTts} case, a choice of radiative variables has
been made which limits the explicit $k^\mu$ dependence in $A'()$ to the
invariant mass arguments. We have defined $u_{14}\equiv(p_1-p_4)^2$
and $u_{23}\equiv(p_2-p_3)^2$ in the above. To arrive at the
corresponding soft photon amplitude one follows an analogous procedure
to that used for the {\sc TsTts} amplitude---the result is
\begin{eqnarray}
{\cal M}_{\mbox{\small\sc TuTts}}\cdot\epsilon & = & 
eQ_3\left(\frac{p_3\cdot\epsilon}{p_3\cdot k} 
-\frac{(p_2-p_3)\cdot\epsilon}{(p_2-p_3)\cdot k}\right)
A'\left(u_{14},t_{24}; 
m_1^2, m_2^2, m_3^2, m_4^2\right) \nonumber \\ & & + 
eQ_4\left(\frac{p_4\cdot\epsilon}{p_4\cdot k} 
-\frac{(p_1-p_4)\cdot\epsilon}{(p_1-p_4)\cdot k}\right)
A'\left(u_{23},t_{13}; 
m_1^2, m_2^2, m_3^2, m_4^2\right) \nonumber \\ & & -
eQ_1\left(\frac{p_1\cdot\epsilon}{p_1\cdot k} 
-\frac{(p_1-p_4)\cdot\epsilon}{(p_1-p_4)\cdot k}\right)
A'\left(u_{23},t_{24}; 
m_1^2, m_2^2, m_3^2, m_4^2\right) \nonumber \\ & & -
eQ_2\left(\frac{p_2\cdot\epsilon}{p_2\cdot k} 
-\frac{(p_2-p_3)\cdot\epsilon}{(p_2-p_3)\cdot k}\right)
A'\left(u_{14},t_{13}; 
m_1^2, m_2^2, m_3^2, m_4^2\right).
\label{eq:tuttssoft}
\end{eqnarray}
During the derivation the constraint $Q_1=Q_3$, $Q_2=Q_4$ once again
arises when we disallow unphysical contributions to the internal 
radiation part of the amplitude. 

In sections~\ref{sectIII} and~\ref{sectIV} we shall consider certain
problems which arise in the application of soft photon
amplitudes. The expressions derived in this section---the
Low-$(\overline{s},\overline{t})$ amplitude of Eq.~(\ref{eq:lowsoft}), 
the {\sc TsTts} amplitude of Eq.~(\ref{eq:tsttssoft}) and the
{\sc TuTts} amplitude of Eq.~(\ref{eq:tuttssoft})---will serve as
instructive examples which demonstrate how these problems arise and in
which circumstances they may be avoided.

\section{Phase Space Problem}
\label{sectIII}

The soft photon approximation is useful in that it provides a relatively 
simple link between the low energy part of a measured photon spectrum 
and the measured cross section for the corresponding non-radiative 
process. 

It is therefore reasonable to insist that a useful soft photon amplitude
must not require evaluations of the non-radiative cross section at 
unphysical, unmeasurable points. Unfortunately, as we shall 
show in this section, this condition is not satisfied by certain soft photon
theorems in the literature. Whether the condition is upheld or not depends 
both upon the choice of radiative phase space variables one uses to 
parameterize the non-radiative amplitude during the construction of
the soft photon amplitude, and upon the the masses of the
particles involved in the scattering. 

The crucial step in the construction of a soft photon amplitude is the 
expansion of any off-mass-shell non-radiative amplitudes about points 
where the kinematic variables have had all explicit dependence on 
photon momentum $k^\mu$ removed, i.e. $k^\mu$ has been set to zero 
wherever it appears.
We will show that even after such an expansion the value of the
non-radiative amplitude may still be required at points outside of the 
region where it is measurable by experiment.

For example, a particular off-shell non-radiative amplitude appearing in the 
derivation of the {\sc TsTts} soft photon amplitude of 
Eq.~(\ref{eq:tsttsextern}) is 
\begin{eqnarray}
A(s_{34},t_{24};m_1^2-2k\cdot p_1,m_2^2,m_3^2,m_4^2) \nonumber 
\end{eqnarray}
where we are considering radiation from particle $1$. 
The soft photon prescription states that
we make a Taylor series expansion about the point where explicit dependence on
$k^\mu$ has been set to zero. For our example this point would be
\begin{eqnarray}
A(s_{34},t_{24};m_1^2,m_2^2,m_3^2,m_4^2). \nonumber 
\end{eqnarray}
This point can be termed on-shell because it is evaluated with 
$p_1^2 = m_1^2$. However, the function $A(s,t;m_1^2,m_2^2,m_3^2,m_4^2)$
is only physically measurable within the region of the $(s,t)$ plane
defined by non-radiative kinematics. We have no guarantee that the region 
$(s_{34},t_{24})$ obtained by evaluating $s_{34},t_{24}$ at values of 
the $p_i$ satisfying radiative phase space constraints is 
contained within this measurable area. 
Indeed, for most choices of radiative variable pairs and for most sets
of masses $m_1,m_2,m_3,m_4$ defining phase space, we find that the soft 
photon amplitude does indeed require evaluations of the non-radiative
amplitude at points which are not physically measurable. 

Since the arguments of this section will depend only on kinematic
constraints and not on the spin structure of the scattering process,
we employ the spinless formalism of the previous section though we
shall be discussing the kinematics applicable to the interactions $\pi^-
p\rightarrow\pi^- p\gamma$ and $pp\rightarrow pp\gamma$. 

\begin{figure} 
\unitlength1.0cm
\begin{center}
\begin{picture}(14.0,7.0) 
\put(2,0){\epsfig{file=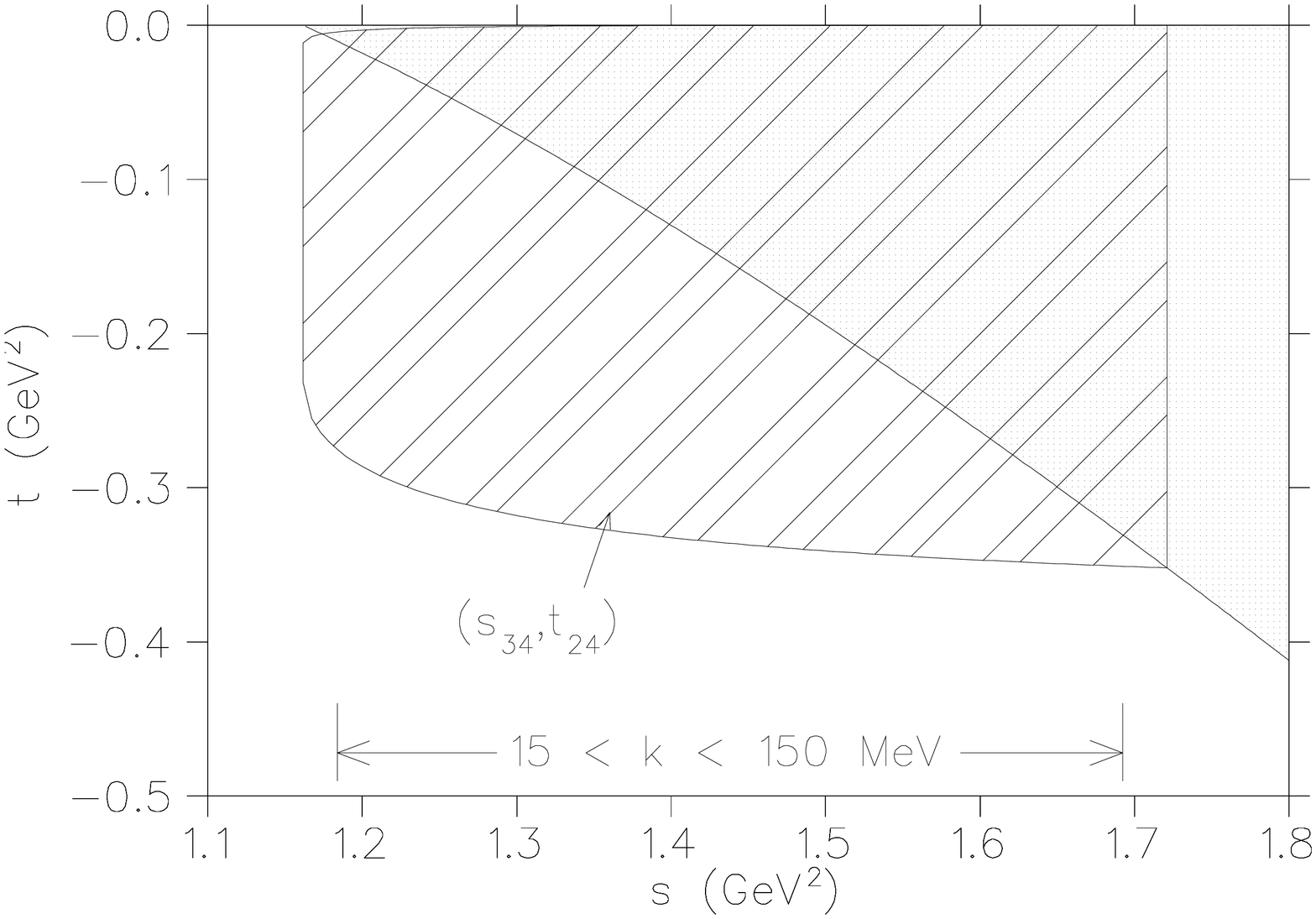,angle=90,height=7cm}}
\end{picture} 
\end{center}
\caption{The shaded region shows the physical region of phase space 
for the elastic process of $\pi^-p$ scattering at pion beam kinetic
energy 298~MeV (i.e. $m_1 = m_3 = m_{\pi^-}$, $m_2 = m_4 =
m_{\mbox{\small proton}}$). The hashed area shows the region 
covered by the radiative phase space point $(s_{34},t_{24})$. 
This area extends far outside of the physical elastic region. The
range of $s_{34}$ marked $15<k<150$~MeV is the approximate region of
radiative phase space studied in the $\pi^-p\gamma$ experiment of
Ref.~\protect\cite{nefkens78}.} 
\label{fig:pip}
\end{figure}

We begin with a simple example, choosing the masses
$m_1=m_3=m_{\pi^-}$ and $m_2=m_4=m_p$, and considering the {\sc TsTts}
soft photon amplitude of Eq.~(\ref{eq:tsttssoft}) for a laboratory pion energy
of 298 MeV, corresponding to a typical experiment \cite{nefkens78}. The 
shaded region
of Fig.~\ref{fig:pip} shows the physically accessible part of the
$(s,t)$ plane---the amplitude $A(s,t)$ would be known over this region
if the elastic process $\pi^-p\rightarrow\pi^-p$ had been
measured at all scattering angles and for interaction energies up to 
$\sqrt{s}\approx1.35$~GeV. The {\sc TsTts} soft photon amplitude 
calls for the evaluation of the
non-radiative amplitude at four points, one of these being
$A(s_{34},t_{24})$. In order to calculate this soft photon amplitude
for initial state pion kinetic energy of 298~MeV in the laboratory
frame and for all allowed energies and orientations of the final state
particles, it turns out that we require the function $A()$ for values of 
$s_{34},t_{24}$ corresponding to the
hashed region shown in Fig.~\ref{fig:pip}. This clearly extends far
outside of the region where the non-radiative amplitude $A()$ is
measurable. Thus, for certain kinematics, the {\sc TsTts} soft photon
approximation to the bremsstrahlung amplitude will not be calculable
unless one is prepared to make a model dependent extrapolation of the
non-radiative amplitude $A()$ outside of its measurable region. The 
introduction of
any such model dependence would remove the usefulness of the soft
photon approximation as an unambiguous method of relating the
$\pi^-p\rightarrow \pi^-p$ process to the radiative process 
$\pi^-p\rightarrow \pi^-p\gamma$. 
The evaluation point $(s_{34},t_{13})$ also suffers from this
problem. The remaining two points in the {\sc TsTts} amplitude,
$(s_{12},t_{13})$ and $(s_{12},t_{24})$, may be shown to lie inside
the measurable region of non-radiative phase space for any elastic
scattering process. 

We can see intuitively how this problem arises. The quantities $s_{12}$ and
$s_{34}$ are related by 
\begin{eqnarray}
  s_{34} = s_{12} - k\cdot (p_1+p_2+p_3+p_4). \nonumber
\end{eqnarray}
As photon energy increases $s_{34}$ becomes progressively smaller than 
$s_{12}$. Even though a range $(s_{12},t_{24})$ as defined by radiative
kinematics might lie within the non-radiative region for $s=s_{12}$, if we
take $s=s_{34}$ we find the allowed range of the non-radiative variable $t$
to be much smaller. The points $(s_{34},t_{24})$ may not be contained within 
this non-radiative physical region. 

\begin{figure} 
\unitlength1.0cm
\begin{center}
\begin{picture}(14.0,15) 
\put(2,7.5){\epsfig{file=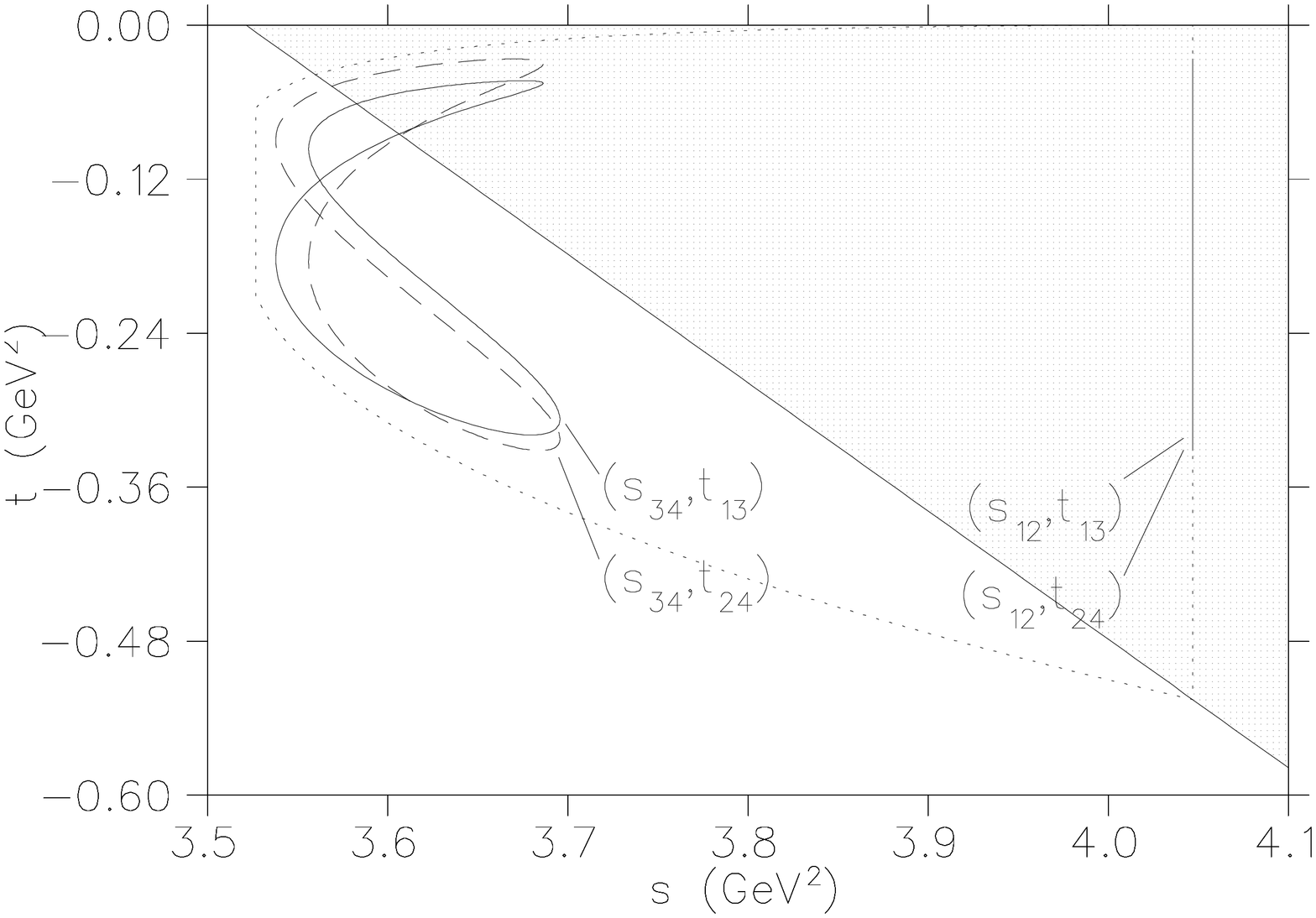,angle=90,height=7cm}}
\put(2,0){\epsfig{file=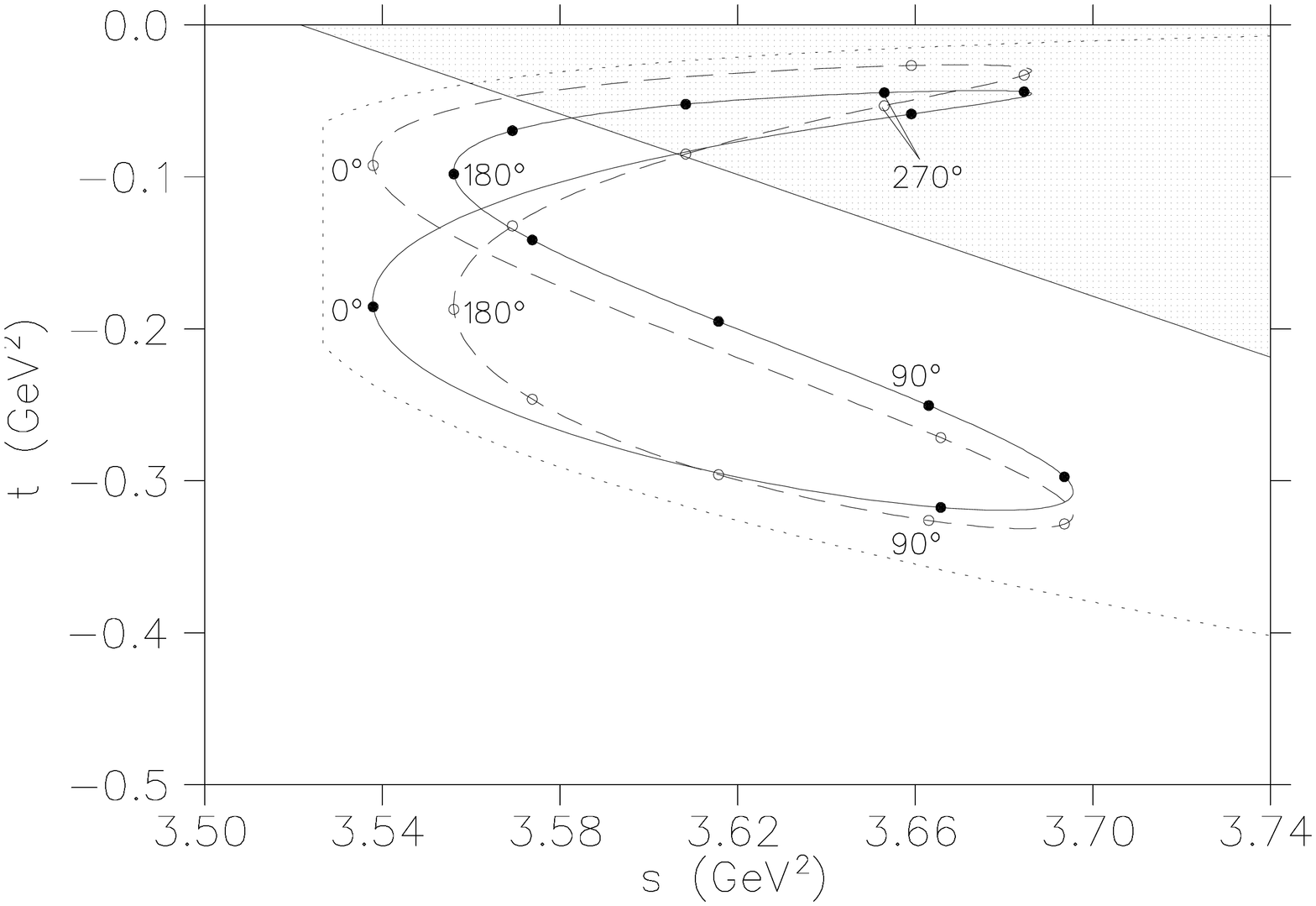,angle=90,height=7cm}}
\end{picture} 
\end{center}
\caption{The shaded area is the physical region of non-radiative phase
space for proton-proton elastic scattering. The area enclosed by
the dotted line is the region mapped out by the points
$(s_{34},t_{13})$ and $(s_{34},t_{24})$ for the corresponding
bremsstrahlung process, operating at proton beam energy of
280~MeV---this corresponds to the beam energy in the experiment of
Ref.~\protect\cite{michaelian90}. The particular regions of radiative
phase space studied in that experiment are also shown---to reproduce 
the experiment's kinematics we fix the outgoing protons at angles of 
$12.4^\circ$ and $12^\circ$ to the beamline in the laboratory frame. 
The points $(s_{34},t_{13})$ and $(s_{34},t_{24})$ are seen
to be outside of the measurable region of elastic phase space for most
photon angles. The lower plot is an expansion of the upper and shows 
the photon angle measured in the target frame for the regions 
$(s_{34},t_{13})$ and $(s_{34},t_{24})$. To guide the eye points have been 
marked at $30^\circ$ intervals in photon lab angle along these trajectories.}
\label{fig:pp}
\end{figure}

This problem is by no means isolated to the one example shown above. For
the case of proton-proton scattering we can also make the same comparison   
between the physical elastic region of phase space and the regions mapped 
out by the radiative variable pairs needed to evaluate the {\sc TsTts} 
soft photon amplitude. The results of this comparison
are shown in Fig.~\ref{fig:pp} for a typical set of kinematics corresponding to
the TRIUMF experiment \cite{michaelian90}. In this case also, the parts of the 
soft photon
amplitude employing the expansion points $A(s_{12}, t_{13})$ or 
$A(s_{12}, t_{24})$ would require only measurable information about the 
non-radiative, elastic amplitude. The parts of the radiative amplitude 
using the points
$A(s_{34}, t_{13})$ or $A(s_{34}, t_{24})$ would, however, require 
unphysical information and would be incalculable unless one resorted to
model-dependent extrapolations of the elastic amplitude.

This difficulty might be avoided by considering only certain experimental
kinematics for the bremsstrahlung process. This is clearly unsatisfactory,
however, particularly when we note that most modern experiments cover
kinematic ranges for which the points $(s_{34}, t_{13})$ 
or $(s_{34}, t_{24})$ lie outside the region accessible in the 
non-radiative process. For example
the $\pi^-p\gamma$ experiment of Ref.~\cite{nefkens78} covered the
majority of radiative phase space, with photon energy in the range 
$15$--$150$~MeV being measured. This region is shown in
Fig.~\ref{fig:pip}. 
The 200~MeV $pp$ bremsstrahlung 
experiment of Ref.~\cite{rogers80} measured the 
photon spectrum as a function of angle with outgoing proton angles fixed 
at $16.4^\circ$ on either side of the beam axis in the lab frame and with 
all particles coplanar. For these kinematics the result is analogous to that
of Fig.~\ref{fig:pp} with the resulting 
trajectories through radiative phase 
space of the points $A(s_{34}, t_{13})$ and $A(s_{34}, t_{24})$ falling 
outside of the physical region of phase space for the elastic $pp$ process.

In contrast to the {\sc TsTts} soft photon amplitude 
the Low-$(\overline{s},\overline{t})$ soft photon amplitude of
Eq.~(\ref{eq:lowsoft}) relies on a single evaluation of the
non-radiative amplitude, at $A(\overline{s},\overline{t})$. For
elastic scattering processes such as $\pi^-p\rightarrow\pi^-p$ our
numeric studies have found that the physical region of the radiative
variables $(\overline{s},\overline{t})$ can fall slightly outside of
the measurable non-radiative region of phase space. For practical
purposes, however, only a very tiny region of radiative phase space
must be excluded in one's model-independent calculation of the
bremsstrahlung process when using the
Low-$(\overline{s},\overline{t})$ amplitude. 

For the special case of identical particle scattering the radiative 
$(\overline{s},\overline{t})$ region is entirely contained within the
physical non-radiative $(s,t)$ region. This is due 
to the fact that the Mandelstam variables $\overline{s}, 
\overline{t}, \overline{u}$ of a radiative
identical particle scattering process satisfy the same phase space 
constraints as the $s, t, u$ of the 
corresponding non-radiative process. 
For the non-radiative process we have the familiar constraints for
equal mass, two-body  elastic scattering
\begin{eqnarray}
s+t+u & = & 4m^2 \nonumber \\
s & \geq & 4m^2 \nonumber \\
0 \geq t,u & \geq & -(s-4m^2).
\end{eqnarray}
For the radiative process we can use four-momentum conservation to write 
\begin{eqnarray}
(k^\mu)^2 & = & (p_1^\mu + p_2^\mu - p_3^\mu - p_4^\mu)^2 \nonumber \\
\Longrightarrow\quad\overline{s}+\overline{t}+\overline{u} & = & 4m^2
\label{eq:radcon}
\end{eqnarray}
where $\overline{u}\equiv\mbox{\small${1\over 2}$}(p_1-p_4)^2+
\mbox{\small${1\over 2}$}(p_2-p_3)^2$.
We also have the constraints,
\begin{eqnarray}
\overline{s} & \geq & 4m^2 \nonumber \\
0 \geq \overline{t},\overline{u} & \geq & -(\overline{s}-4m^2).
\label{eq:radcon2}
\end{eqnarray}
The threshold condition on $\overline{s}$ is clear, however the $\overline{t},
\overline{u}$ constraints require some explanation. 
For identical particle scattering it may be shown, by considering the
appropriate rest frames, that the variables
$t_{13}$, and symmetrically $t_{24}$, have zero as their upper
bounds. Thus the average $\overline{t} = \frac{1}{2}(t_{13}+t_{24})$ 
is also bounded above by zero. Putting $\overline{t} \leq 0$ 
into Eq.~(\ref{eq:radcon}) 
we find the lower bound for $\overline{u}$; $\overline{u} 
\geq -(\overline{s}-4m^2)$. Finally, noting 
that the constraints on $\overline{t}$ and on $\overline{u}$ must be 
the same for identical particle scattering through the symmetry of the
kinematics under the interchange of final state particles, we have the result 
of Eq.~(\ref{eq:radcon2}). Our conclusion is that for identical
particle scattering, all points $(\overline{s},\overline{t})$ defined
by radiative phase space constraints lie within the allowed $(s,t)$
region of the corresponding non-radiative process. Numeric studies of
the phase space regions confirm this conclusion. 

From numeric studies of the kinematics for the interactions 
$\pi^-p\rightarrow\pi^-p$, $pp\rightarrow pp$ and the corresponding
bremsstrahlung processes, it appears that the points used in the 
{\sc TuTts} soft photon amplitude lie inside the measurable
non-radiative region, at least for the kinematic conditions relevant to
existing experiments. This implies that, for these interactions, the 
{\sc TuTts} amplitude does not suffer from the phase space problem.

\section{Symmetrization Problem}
\label{sectIV}

In section~\ref{sectV} we will show that 
a problem exists with correctly antisymmetrizing 
the spin-$\frac{1}{2}$ {\sc TsTts} and {\sc TuTts} amplitudes of 
Ref.~\cite{liou95}. 
To illustrate the source of this problem we consider
in this section the quite analogous and algebraically simpler symmetrization 
of the spin-$0$ {\sc TsTts} and {\sc TuTts} amplitudes.

The spin-$0$ amplitudes given in section~\ref{sectII} would have to be 
explicitly symmetrized if applied to the case of identical particle scattering.
Upon attempting to symmetrize the {\sc TsTts} and {\sc TuTts} amplitudes
we find that the connection to measurable non-radiative scattering data is
lost. More specifically, the symmetrized radiative {\sc TsTts} and 
{\sc TuTts} amplitudes cannot be written in terms of the symmetrized 
non-radiative amplitudes, and thus cannot be evaluated directly from 
experimental information on the non-radiative process. 
This problem is quite independent of the phase space problem discussed
previously. The Low amplitude, employing a single choice of Taylor expansion 
point at $(\overline{s},\overline{t})$, is also treated for comparison. It
is found that the symmetrized $(\overline{s},\overline{t})$ soft photon
amplitude may be written in terms of the measurable, symmetrized non-radiative
scattering amplitude. This is due to a special property of the $(\overline{s}$,
$\overline{t})$ variables. 

We denote the unsymmetrized non-radiative scattering amplitude by $A(s,t)$, 
where we have suppressed the invariant mass arguments of this function. 
The symmetrized amplitude, which we obtain by adding in the amplitude with 
$p_3 \leftrightarrow p_4$, is then 
\begin{eqnarray}
A^S(s,t) & \equiv & A(s,t) + A(s,u) \nonumber \\
 & = & A(s,t) + A(s,4m^2-s-t).
\label{eq:Asym}
\end{eqnarray}
The unsymmetrized Low-$(\overline{s},\overline{t})$ amplitude may be written 
\cite{fearing72b}
\begin{eqnarray}
{\cal M}_{(\overline{s},\overline{t})}\cdot\epsilon & = & 
eQ\left\{\left[\frac{p_3\cdot\epsilon}{p_3\cdot k}+{\cal D}^\mu(p_3) 
\frac{\partial}{\partial p_3^\mu} \right] A(\overline{s},\overline{t}) \right.
+ \left[ \frac{p_4\cdot  \epsilon}{p_4\cdot  k} + {\cal D}^\mu(p_4) 
\frac{\partial}{\partial p_4^\mu} \right] A(\overline{s},\overline{t}) 
\nonumber \\
&&\hspace{4.5mm}-\left[\frac{p_1\cdot\epsilon}{p_1\cdot k}-{\cal D}^\mu(p_1) 
\frac{\partial}{\partial p_1^\mu} \right] A(\overline{s},\overline{t}) 
\left.-\left[\frac{p_2\cdot\epsilon}{p_2\cdot k}-{\cal D}^\mu(p_2) 
\frac{\partial}{\partial p_2^\mu} \right] A(\overline{s},\overline{t}) \right\}
\label{eq:unsym_low} 
\end{eqnarray}
where, following Ref.~\cite{fearing72b}, we introduce the notation
\begin{eqnarray}
{\cal D}^\mu(p)&\equiv&\frac{p\cdot\epsilon}{p\cdot k}k^\mu 
- \epsilon^\mu. 
\end{eqnarray}
This is easily related to the usual form of Eq.~(\ref{eq:lowsoft}) 
for the amplitude by noting that
\begin{eqnarray}
\frac{\partial}{\partial p_1^\mu} = 
\frac{\partial \overline{s}}{\partial p_1^\mu} 
\frac{\partial}{\partial \overline{s}} + 
\frac{\partial \overline{t}}{\partial p_1^\mu} 
\frac{\partial}{\partial \overline{t}} = 
(p_{1\mu}+p_{2\mu}) \frac{\partial}{\partial \overline{s}} +
(p_{1\mu}-p_{3\mu}) \frac{\partial}{\partial \overline{t}} 
\end{eqnarray}
with similar expressions for the other derivatives. 

The symmetrized amplitude is then ${\cal M}^{S}\cdot\epsilon \equiv 
{\cal M}_{(12\rightarrow34)}\cdot\epsilon + 
{\cal M}_{(12\rightarrow43)}\cdot\epsilon$,
\begin{eqnarray}
{\cal M}^{S}_{(\overline{s},\overline{t})}\cdot\epsilon & = & 
eQ \left\{ \left[ \frac{p_3\cdot  \epsilon}{p_3\cdot  k} + {\cal D}^\mu(p_3) 
\frac{\partial}{\partial p_3^\mu} \right] A(\overline{s},\overline{t}) +
\left[ \frac{p_4\cdot  \epsilon}{p_4\cdot  k} + {\cal D}^\mu(p_4) 
\frac{\partial}{\partial p_4^\mu} \right] A(\overline{s},\overline{u}) \right.
\nonumber \\
& & \hspace{4.5mm}
+ \left[ \frac{p_4\cdot  \epsilon}{p_4\cdot  k} + {\cal D}^\mu(p_4) 
\frac{\partial}{\partial p_4^\mu} \right] A(\overline{s},\overline{t}) 
+ \left[ \frac{p_3\cdot  \epsilon}{p_3\cdot  k} + {\cal D}^\mu(p_3) 
\frac{\partial}{\partial p_3^\mu} \right] A(\overline{s},\overline{u}) 
\nonumber \\
& & \hspace{4.5mm}
- \left[ \frac{p_1\cdot  \epsilon}{p_1\cdot  k} - {\cal D}^\mu(p_1) 
\frac{\partial}{\partial p_1^\mu} \right] A(\overline{s},\overline{t}) 
- \left[ \frac{p_1\cdot  \epsilon}{p_1\cdot  k} - {\cal D}^\mu(p_1) 
\frac{\partial}{\partial p_1^\mu} \right] A(\overline{s},\overline{u}) 
\nonumber \\
& & \hspace{4.5mm}
- \left. \left[ \frac{p_2\cdot  \epsilon}{p_2\cdot  k} - {\cal D}^\mu(p_2) 
\frac{\partial}{\partial p_2^\mu} \right] A(\overline{s},\overline{t}) 
- \left[ \frac{p_2\cdot  \epsilon}{p_2\cdot  k} - {\cal D}^\mu(p_2) 
\frac{\partial}{\partial p_2^\mu} \right] A(\overline{s},\overline{u}) 
\right\} . 
\label{eq:4.50}
\end{eqnarray}
A common factor in this expression is 
\[
A(\overline{s}, \overline{t}) + A(\overline{s}, \overline{u}).
\]
In order to have the soft photon amplitude solely a function of the
measurable, symmetric part of the non-radiative amplitude we must be
able to write this factor in terms of the symmetric function $A^S()$.
From Eq.~(\ref{eq:Asym}) it is clear that since the relation 
$\overline{s}+\overline{t}+\overline{u}=4m^2$ holds among the radiative
variables we can write 
\begin{eqnarray}
A(\overline{s}, \overline{t}) + A(\overline{s}, \overline{u}) & = & 
A(\overline{s}, \overline{t}) + A(\overline{s}, 
4m^2-\overline{s}-\overline{t}) \nonumber \\
& = & A^S(\overline{s}, \overline{t}). 
\end{eqnarray}
The symmetrized radiative amplitude now takes on the form of the 
unsymmetrized amplitude, but with $A(\overline{s}, \overline{t})$ 
replaced by the measurable, symmetrized non-radiative amplitude 
$A^S(\overline{s}, \overline{t})$. 
\begin{eqnarray}
{\cal M}^S_{(\overline{s},\overline{t})}\cdot\epsilon & = & 
eQ \left\{ \left[ \frac{p_3\cdot \epsilon}{p_3\cdot k} + {\cal D}^\mu(p_3) 
\frac{\partial}{\partial p_3^\mu} \right] 
A^S(\overline{s},\overline{t}) \right. 
+ \left[ \frac{p_4\cdot \epsilon}{p_4\cdot k} + {\cal D}^\mu(p_4) 
\frac{\partial}{\partial p_4^\mu} \right] 
A^S(\overline{s},\overline{t}) \nonumber \\
&& \hspace{4.5mm}
- \left[ \frac{p_1\cdot \epsilon}{p_1\cdot k} - {\cal D}^\mu(p_1) 
\frac{\partial}{\partial p_1^\mu} \right] 
A^S(\overline{s},\overline{t})
- \left. \left[ \frac{p_2\cdot \epsilon}{p_2\cdot k} - {\cal D}^\mu(p_2) 
\frac{\partial}{\partial p_2^\mu} \right] 
A^S(\overline{s},\overline{t}) \right\}  
\end{eqnarray}

This procedure of correctly symmetrizing the radiative amplitude by simply
replacing $A() \rightarrow A^S()$ works only for the 
Low-$(\overline{s},\overline{t})$ case, due to the  
relationship $\overline{s}+\overline{t}+\overline{u}=4m^2$ which holds only for
this specific Low choice of variables. We will now show 
explicitly that such a replacement in the {\sc TsTts} or {\sc TuTts}
amplitudes does not work and that these amplitudes cannot be expressed in terms
of symmetrized non-radiative amplitudes.

For the radiative process, the unsymmetrized {\sc TsTts} amplitude of 
Eq.~(\ref{eq:tsttssoft}) is 
\begin{eqnarray}
{\cal M}_{\mbox{\small\sc TsTts}}^\mu & = & 
eQ\left(\left[\frac{p_3^\mu}{p_3\cdot k}-\frac{(p_3+p_4)^\mu}
{(p_3+p_4)\cdot k}\right]A\left(s_{12},t_{24}\right)\right.
+\left[\frac{p_4^\mu}{p_4\cdot k}-\frac{(p_3+p_4)^\mu}{(p_3+p_4)\cdot k}
\right]A\left(s_{12},t_{13}\right)\nonumber\\&&\hspace{4mm}
-\left[\frac{p_1^\mu}{p_1\cdot k}-\frac{(p_1+p_2)^\mu}{(p_1+p_2)\cdot k}
\right]A\left(s_{34},t_{24}\right)-\left.\left[\frac{p_2^\mu}{p_2\cdot k}
-\frac{(p_1+p_2)^\mu}{(p_1+p_2)\cdot k}\right]
A\left( s_{34}, t_{13} \right) \right) . 
\label{eq:unsym_tstts} 
\end{eqnarray}
We define the symmetrized amplitude ${\cal M}_
{\mbox{\small\sc TsTts}}^{S\mu} \equiv 
{\cal M}^\mu_{(12\rightarrow34)} + {\cal M}^\mu_{(12\rightarrow43)}$,
\begin{eqnarray}
{\cal M}_{\mbox{\small\sc TsTts}}^{S\mu} & = & 
eQ \left( \left[ \frac{p_3^\mu}{p_3\cdot k} - \frac{(p_3+p_4)^\mu}
{(p_3+p_4)\cdot k} 
\right] \left( A\left( s_{12}, t_{24} \right) + A\left( s_{12}, u_{14} \right) 
\right) \right. \nonumber \\ & & \hspace{2.5mm} + 
\left[ \frac{p_4^\mu}{p_4\cdot k} - \frac{(p_3+p_4)^\mu}{(p_3+p_4)\cdot k} 
\right] 
\left( A\left( s_{12}, t_{13} \right) + A\left( s_{12}, u_{23} \right) 
\right) \nonumber \\ & & \hspace{2.5mm} -
\left[ \frac{p_1^\mu}{p_1\cdot k} - \frac{(p_1+p_2)^\mu}{(p_1+p_2)\cdot k} 
\right] 
\left( A\left( s_{34}, t_{24} \right) +A\left( s_{34}, u_{23} \right) 
\right) \nonumber \\ & & \hspace{2.5mm} - 
\left. \left[ \frac{p_2^\mu}{p_2\cdot k} - \frac{(p_1+p_2)^\mu}{(p_1+p_2)
\cdot k} 
\right] \left( A\left( s_{34}, t_{13} \right) + A\left( s_{34}, u_{14} \right) 
\right) \right) .
\label{eq:sym_tstts}
\end{eqnarray}
We define symmetrized functions in analogy to the non-radiative case:
\begin{eqnarray}
A^{S_1}(s_{34},t_{24},u_{23}) & \equiv & A(s_{34},t_{24}) + A(s_{34},u_{23})
\nonumber \\
A^{S_2}(s_{34},t_{13},u_{14}) & \equiv & A(s_{34},t_{13}) + A(s_{34},u_{14})
\nonumber \\
A^{S_3}(s_{12},t_{24},u_{14}) & \equiv & A(s_{12},t_{24}) + A(s_{12},u_{14})
\nonumber \\
A^{S_4}(s_{12},t_{13},u_{23}) & \equiv & A(s_{12},t_{13}) + A(s_{12},u_{23}).
\label{eq:Asym_tstts}
\end{eqnarray}
Notice that these functions are not the same as the symmetric non-radiative
function $A^S(s,t)$ which would be measured in the non-radiative process. 
This is because an internal constraint similar to that for the non-radiative 
phase space variables, $s+t+u=4m^2$, 
does not hold for $s_{34}$, $t_{24}$ and $u_{23}$, or for the other sets 
of radiative variables which appear as arguments in the $A^{S_i}$. Instead one
has relations like $s_{34}+t_{24}+u_{23} = 4m^2-2k\cdot p_1$. Thus direct 
replacement of, for example, $A^{S_1}(s_{34},t_{24},u_{23})$ by 
$A^S(s_{34},t_{24})$ will give an error of the form 
\begin{eqnarray}
A^{S_1}(s_{34},t_{24},u_{23})-A^S(s_{34},t_{24})&=&
-2p_1\cdot k{\partial\over\partial u}
\left.A(s_{34},u)\right|_{u=u_{23}}+{\cal O}(k^2)
\end{eqnarray}
One would naively 
expect this ${\cal O}(k)$ error which is being introduced to give rise to an
${\cal O}(k/k)$ or ${\cal O}(1)$ error in the amplitude 
${\cal M}_{\mbox{\small\sc TsTts}}^{S\mu}$. Due to a cancellation in this 
leading order between the four $A^{S_i}()$ terms appearing in 
${\cal M}_{\mbox{\small\sc TsTts}}^{S\mu}$, the error introduced by these
replacements is instead only of ${\cal O}(k)$. Thus in this case the error is 
of the same order as other terms dropped in the derivation of the amplitude.

For the {\sc TuTts} amplitude another difficulty arises in trying to
symmetrize. The unsymmetrized amplitude is given by Eq.~(\ref{eq:tuttssoft}) 
\begin{eqnarray}
{\cal M}_{\mbox{\small\sc TuTts}}^\mu & = & 
eQ\left(\left[\frac{p_3^\mu}{p_3\cdot k}-\frac{(p_2-p_3)^\mu}
{(p_2-p_3)\cdot k}\right]A'\left( u_{14}, t_{24} \right) \right.
+\left[\frac{p_4^\mu}{p_4\cdot k}-\frac{(p_1-p_4)^\mu}
{(p_1-p_4)\cdot k}\right]A'\left(u_{23},t_{13}\right)\nonumber\\
&&\hspace{4.5mm}-\left[\frac{p_1^\mu}{p_1\cdot k}
-\frac{(p_1-p_4)^\mu}{(p_1-p_4)\cdot k}\right] 
A'\left( u_{23}, t_{24} \right)-\left.\left[\frac{p_2^\mu}{p_2\cdot k}
-\frac{(p_2-p_3)^\mu}{(p_2-p_3)\cdot k} 
\right] A'\left( u_{14}, t_{13} \right) \right) . 
\label{eq:unsym_tutts}  
\end{eqnarray}
and we define the symmetrized amplitude as ${\cal M}^{S\mu}_
{\mbox{\small\sc TuTts}} \equiv 
{\cal M}^\mu_{(1,2\rightarrow 3,4)} + {\cal M}^\mu_{(1,2\rightarrow 4,3)}$, 
so that
\begin{eqnarray}
{\cal M}_{\mbox{\small\sc TuTts}}^{S\mu} & = &    
eQ \left( \left[ \frac{p_3^\mu}{p_3\cdot k} - \frac{(p_2-p_3)^\mu}
{(p_2-p_3)\cdot k} 
\right] A'\left( u_{14}, t_{24} \right) + \left[ \frac{p_3^\mu}{p_3\cdot k} - 
\frac{(p_1-p_3)^\mu}{(p_1-p_3)\cdot k} \right] A'\left( t_{24}, u_{14} \right) 
\right. \nonumber \\ && \hspace{4.5mm} + 
\left[ \frac{p_4^\mu}{p_4\cdot k} - \frac{(p_1-p_4)^\mu}{(p_1-p_4)\cdot k} 
\right] 
A'\left( u_{23}, t_{13} \right) + \left[ \frac{p_4^\mu}{p_4\cdot k} - 
\frac{(p_2-p_4)^\mu}{(p_2-p_4)\cdot k} \right] A'\left( t_{13}, u_{23} \right) 
\nonumber \\ && \hspace{4.5mm}  -
\left[ \frac{p_1^\mu}{p_1\cdot k} - \frac{(p_1-p_4)^\mu}{(p_1-p_4)\cdot k} 
\right] 
A'\left( u_{23}, t_{24} \right) - \left[ \frac{p_1^\mu}{p_1\cdot k} - 
\frac{(p_1-p_3)^\mu}{(p_1-p_3)\cdot k} \right] 
A'\left( t_{24}, u_{23} \right) \nonumber \\  && \hspace{4.5mm} - 
\left. \left[ \frac{p_2^\mu}{p_2\cdot k} - \frac{(p_2-p_3)^\mu}
{(p_2-p_3)\cdot k} 
\right] A'\left( u_{14}, t_{13} \right) - \left[ \frac{p_2^\mu}{p_2\cdot k} - 
\frac{(p_2-p_4)^\mu}{(p_2-p_4)\cdot k} 
\right] A'\left( t_{13}, u_{14} \right) \right) .
\label{eq4:100}
\end{eqnarray}
Consistent with our previous definition of $A'()$ we have, for example,
$A'(t_{24},u_{14})=A(\sum_im_i^2-t_{24}-u_{14},u_{14})$ where $A(s,t)$ is
the non-symmetrized amplitude for the non-radiative process. Again there is
difficulty because the radiative amplitude is not expressed in terms of the
symmetrized non-radiative amplitude, which is all that can be measured.
The terms in the square brackets of the form $p_i/k\cdot p_i$ can in fact be
factored leaving a correctly symmetrized combination of the $A'$ as an overall
factor. However, the problem arises with the $(p_i-p_j)^\mu/ k\cdot (p_i-p_j)$
terms, which were added to make the amplitude gauge invariant. The momenta are
such that these terms cannot be factored leaving just the symmetrized 
amplitude.

The procedure followed in Ref.~\cite{liou95} is to take for 
the symmetrized radiative amplitude the original unsymmetrized {\sc TuTts} 
amplitude with the functions $A'()$ replaced by their counterparts 
from the symmetrized elastic process. This prescription would give the result 
\begin{eqnarray}
&&eQ\left(\left[\frac{p_3^\mu}{p_3\cdot k}
-\frac{(p_2-p_3)^\mu}{(p_2-p_3)\cdot k}\right]
A'^S\left(u_{14},t_{24}\right)\right.+\left[\frac{p_4^\mu}{p_4\cdot k}
-\frac{(p_1-p_4)^\mu}{(p_1-p_4)\cdot k}\right] 
A'^S\left( u_{23}, t_{13} \right) \nonumber \\ & & \hspace{4.5mm} -
\left[\frac{p_1^\mu}{p_1\cdot k}
-\frac{(p_1-p_4)^\mu}{(p_1-p_4)\cdot k}\right] 
A'^S\left( u_{23}, t_{24} \right)
-\left.\left[\frac{p_2^\mu}{p_2\cdot k}
-\frac{(p_2-p_3)^\mu}{(p_2-p_3)\cdot k} 
\right] A'^S\left( u_{14}, t_{13} \right) \right) 
\label{eq4:110}
\end{eqnarray}
where, for example, 
\begin{eqnarray}
A'^S\left( u_{23}, t_{24}\right) & = & 
A'\left( u_{23}, t_{24}\right) + A'\left( t_{24}, u_{23}\right) \nonumber \\
& = & A\left( 4m^2-u_{23}-t_{24}, t_{24}\right) + 
A\left( 4m^2-u_{23}-t_{24}, u_{23}\right) \nonumber \\
& = & A^S\left( 4m^2-u_{23}-t_{24}, t_{24}\right) \label{eq:Asym_tutts} 
\end{eqnarray}
This expresses the result in terms of measurable non-radiative amplitudes, but
is not completely symmetric under the interchange of particle labels 3 and 4
because of the kinematic factors multiplying $A^S$.

A detailed comparison between the symmetrized amplitude of 
Eq.~(\ref{eq4:100}) and the
form of Eq.~(\ref{eq4:110}) shows them to be unequal. They differ in this case 
by terms of ${\cal O}(k/k)$. Since the  ${\cal O}(k/k)$ terms are uniquely
determined in a soft photon approach and can come only from the diagrams with
radiation from external legs \cite{adler66}, this must mean that this
prescription in effect adds in some  ${\cal O}(k/k)$ terms which are not
allowed by the analyticity requirements of the soft photon approach.
We have found no way to express the correct expression, Eq.~(\ref{eq4:100}), 
solely in terms of the measurable symmetrized non-radiative amplitude,
$A^S(s,t)$, other than simply expanding about $\overline{u}, \overline{t}$ and
thus recovering the original Low amplitude Eq.~(\ref{eq:lowsoft}), up 
to corrections of ${\cal O}(k)$.

Since the problem arises with the $(p_i-p_j)^\mu/ k\cdot (p_i-p_j)$
terms which were added to make the amplitude gauge invariant, an alternative
approach would be to try to find some different gauge term to add which does
not suffer from these problems \cite{LTGprivcom}. To explore this 
possibility rewrite
Eq.~(\ref{eq:unsym_tutts}) as
\begin{eqnarray}
\overline{\cal M}_{\mbox{\small\sc TuTts}}^\mu & = & 
eQ\left(\left[\frac{p_3^\mu}{p_3\cdot k}-\Delta _3^\mu
\right]A'\left( u_{14}, t_{24} \right) \right.
+\left[\frac{p_4^\mu}{p_4\cdot k}-\Delta _4^\mu
\right]A'\left(u_{23},t_{13}\right)\nonumber\\
&&\hspace{4.5mm}-\left[\frac{p_1^\mu}{p_1\cdot k}
-\Delta _1^\mu\right] 
A'\left( u_{23}, t_{24} \right)-\left.\left[\frac{p_2^\mu}{p_2\cdot k}
-\Delta _2^\mu 
\right] A'\left( u_{14}, t_{13} \right) \right) . 
\label{eq:newunsym_tutts}  
\end{eqnarray}
Here the $\Delta _i^\mu$ are structures of the general form $V^\mu/k \cdot V$,
where $V$ is some vector, and originate in the term added to ensure gauge
invariance. In accord with general principles this gauge term must be 
${\cal O}(1)$ and cannot contain terms ${\cal O}(k/k)$. To determine the
conditions imposed on the  $\Delta _i^\mu$ by this analyticity requirement we
expand the $A'()$ about the single point $\overline{s}, \overline{t},
\overline{u}$. The result is that all the $\Delta _i^\mu$, or more precisely,
all $\epsilon \cdot \Delta _i$, must be equal. One can see from
Eq.~(\ref{eq:unsym_tutts}) that for the 
${\cal M}_{\mbox{\small\sc TuTts}}^\mu$ amplitude this is satisfied. 

Next we define $\Delta _i'$ to be $\Delta _i$ with $p_3 \leftrightarrow p_4$ 
and 
symmetrize the amplitude by adding in a piece with $p_3 \leftrightarrow p_4$ as
was done in going from Eq.~(\ref{eq:unsym_tutts}) to
Eq.~(\ref{eq4:100}). Then the requirement that we be able to express the
full symmetrized amplitude in terms of the symmetrized elastic amplitudes of
Eq.~(\ref{eq:Asym_tutts}), which are the measurable quantities, requires that
$\Delta _1 = \Delta _1'$, $\Delta _2 = \Delta _2'$,   $\Delta _3 = 
\Delta _4'$, and
$\Delta _4 = \Delta _3'$. This condition is not satisfied by the
$\Delta _i$ of the symmetrized {\sc TuTts} amplitude of Eq.~(\ref{eq4:100})
and that is the reason that the  {\sc TuTts} amplitude cannot be written in a
correctly symmetrized form. 

Putting these requirements on the $\Delta _i$ together, we find that the
$\Delta _i$ must all be the same and must be symmetric in the interchange    
$p_3 \leftrightarrow p_4$. That is not true for the $\Delta _i$ of the
{\sc TuTts} amplitude, but it is easy to find such $\Delta _i$. For example,
consider the following four $\Delta _i$: 
\begin{eqnarray}
&&\frac{p_3^\mu+p_4^\mu}{k \cdot (p_3+p_4)},  \nonumber  \\
&&\frac{1}{4} \left( \frac{p_1^\mu}{k \cdot p_1}+\frac{p_2^\mu}{k \cdot p_2}
+\frac{p_3^\mu}{k \cdot p_3} + \frac{p_4^\mu}{k \cdot p_4} \right), 
\nonumber \\
&&\frac{1}{2}\left( \frac{p_3^\mu-p_4^\mu}{k \cdot (p_3-p_4)} +
\frac{p_1^\mu-p_2^\mu}{k \cdot (p_1-p_2)}\right), \nonumber \\
&&\frac{1}{2}\left( \frac{p_1^\mu-p_4^\mu}{k \cdot (p_1-p_4)} +
\frac{p_1^\mu-p_3^\mu}{k \cdot (p_1-p_3)}\right).
\end{eqnarray}
Any of these,  when used in the symmetrized version of
Eq.~(\ref{eq:newunsym_tutts}), will produce an amplitude 
$\overline{\cal M}_{\mbox{\small\sc TuTts}}^\mu$  which is gauge invariant, 
satisfies
the analyticity and other requirements of a soft photon theorem, is properly
symmetric in the interchange of identical particles so that only the
measurable symmetrized elastic amplitudes are required, and is expressed in
terms of the same kinematic variables as the original 
${\cal M}_{\mbox{\small\sc TuTts}}^\mu$. All of these amplitudes will be
equivalent, as is generally true of various different soft photon amplitudes,
in the sense that they will differ by terms which are ${\cal O}(k)$.

The arguments above have shown that the {\sc TsTts} and original 
{\sc TuTts} amplitudes
for a spin-$0$ scattering process cannot be made correctly symmetric
under interchange of identical particles while
maintaining the necessary link to the symmetric non-radiative amplitude. 
For the {\sc TsTts} case we found the failure in symmetrization 
to arise at ${\cal O}(k)$ in the amplitude, but for the {\sc TuTts} case it
arises at ${\cal O}(k/k)$ . 
In the next section we extend the derivation of the 
Low-$(\overline{s},\overline{t})$, {\sc TsTts} and {\sc TuTts} amplitudes to
spin-$\frac{1}{2}$ identical particle scattering, and consider the 
calculation of proton-proton bremsstrahlung. We thus show that exactly the same
problems which arose in this section in symmetrizing a spin-0 amplitude arise
also in antisymmetrizing a spin-$\frac{1}{2}$ amplitude.

\section{Extension to Spin-$\mbox{\small${1\over 2}$}$}
\label{sectV}

In this section we will consider the proton-proton bremsstrahlung
process. This requires the extension of
the spinless formalism of section~\ref{sectII} to the scattering of
spin-$\mbox{\small${1\over 2}$}$ particles. Care must also be taken to 
write the $pp$
elastic and the $pp$ bremsstrahlung amplitudes such that they are
antisymmetric under interchange of final state protons.

The unsymmetrized amplitude for elastic proton-proton scattering may be written
\begin{eqnarray}
A[12\rightarrow 34]\equiv \sum_{\alpha=1}^5 F_\alpha(s, t)
\left[ \overline{u}_3 t_\alpha u_1 \right]\left[ \overline{u}_4 t^\alpha u_2 
\right]
\end{eqnarray}
where \[ t_\alpha \equiv \left\{ 1, \mbox{\small${1\over\sqrt{2}}$}
\sigma^{\mu\nu}, 
i\gamma_5\gamma^\mu, \gamma^\mu, \gamma_5 \right\} \] 
with summation over the Lorentz indices of the $t_\alpha$ being implied.
The antisymmetrized $pp$ elastic amplitude is then defined as
\begin{eqnarray}
A^A(s,t)&\equiv&A[12\rightarrow 34]-A[12\rightarrow 43]\nonumber\\
&=&\sum_{\alpha=1}^5F_\alpha(s,t)\left[\overline{u}_3t_\alpha u_1\right]
\left[\overline{u}_4t^\alpha u_2\right]-
\sum_{\alpha=1}^5F_\alpha(s,u)\left[\overline{u}_4t_\alpha u_1\right]
\left[\overline{u}_3t^\alpha u_2\right].
\end{eqnarray}
Using the Fierz relation \cite{fierz55}
\begin{eqnarray}
(t_\alpha)_{\phi\sigma}(t^\alpha)_{\tau\nu} 
& = & \sum_{\beta=1}^5 {\cal C}_{\alpha\beta} 
(t_\beta)_{\phi\nu}(t^\beta)_{\tau\sigma} 
\label{eq:fierzrel}
\end{eqnarray}
with 
\begin{eqnarray}
{\cal C}_{\alpha\beta} & = & \frac{1}{4} \left( 
\begin{array}{rrrrr}
1 & 1 & 1 & 1 & 1 \\
6 & -2 & 0 & 0 & 6 \\
4 & 0 & -2 & 2 & -4 \\
4 & 0 & 2 & -2 & -4 \\
1 & 1 & -1 & -1 & 1 
\end{array}
\right)
\label{eq:fierzmat}
\end{eqnarray}
the antisymmetrized expression $A^A(s,t)$ can be put back into the
form of the unsymmetrized $A[12\rightarrow34]$,
\begin{eqnarray}
A^A(s,t)&=&
\sum_{\alpha=1}^5\left(F_\alpha(s,t)
-\sum_{\beta=1}^5{\cal C}_{\beta\alpha}F_\beta(s,u)\right)
\left[\overline{u}_3t_\alpha u_1\right]
\left[\overline{u}_4t^\alpha u_2\right].
\end{eqnarray}
The correctly antisymmetrized amplitude is then identical to the 
unsymmetrized amplitude but with $F_\alpha(s, t)$ replaced by 
\begin{eqnarray}
F^A_\alpha(s, t) & \equiv & F_\alpha(s, t) - \sum_{\beta=1}^5 
{\cal C}_{\beta \alpha} F_\beta(s, u) 
\label{eq:FAdef}
\end{eqnarray}
where $s+t+u = 4m^2$. It is these functions, $F^A_\alpha(s, t)$, and
not the unsymmetrized $F_\alpha(s, t)$ which are experimentally
accessible through study of $pp$ elastic scattering. We must therefore
ensure that the antisymmetrized forms of our soft photon
approximations to the proton-proton bremsstrahlung amplitude can be 
expressed purely in terms of the $F^A_\alpha(s,t)$, rather than the
$F_\alpha(s,t)$. 

We now consider the form of the three soft photon amplitudes of
section~\ref{sectII} extended to spin-$\mbox{\small${1\over 2}$}$ 
identical particle 
scattering. The unsymmetrized $pp$ bremsstrahlung soft photon
amplitudes may be written in the form 
\begin{eqnarray}
{\cal M}^\mu\epsilon_\mu&=&\sum_{\alpha=1}^5eQ\left[\overline{u}_3
X^\alpha_\mu\epsilon^\mu u_1\overline{u}_4t_\alpha u_2
+\overline{u}_3t_\alpha
u_1\overline{u}_4Y_\mu^\alpha\epsilon^\mu u_2 
\right]
\label{eq:genspinhalf}
\end{eqnarray}
with the functions $X^\alpha_\mu$ and $Y^\alpha_\mu$ taking on a
different form for each of the Low-$(\overline{s},\overline{t})$, the 
{\sc TsTts} and the {\sc TuTts} amplitudes. Using the notation of
Ref.~\cite{liou95} 
\begin{eqnarray}
{\cal R}(p)\cdot\epsilon & = & 
\frac{1}{4} \left[ \not\!\epsilon, \not\! k \right]
+ \frac{i \kappa_p}{8m} \left\{ \left[ \not\!\epsilon, \not\! k \right], 
\not\! p \right\} \nonumber
\end{eqnarray}
where $\kappa=1.79$ and $m$ are the proton anomalous magnetic moment
and mass, we may write the $X^\alpha_\mu$ and $Y^\alpha_\mu$ functions
for each case as follows:
\begin{itemize}
\item 
\begin{flushleft}
Low-$(\overline{s},\overline{t})$ amplitude:
\end{flushleft}
\begin{eqnarray}
X^\alpha\cdot\epsilon & \equiv & 
\left[  
\frac{p_3\cdot\epsilon+{\cal R}(p_3)\cdot\epsilon}{p_3\cdot k} + 
{\cal D}^\mu(p_3) \frac{\partial}
{\partial p_3^\mu} \right] t^\alpha F_\alpha (\overline{s},\overline{t}) 
\nonumber \\
& & - t^\alpha \left[
\frac{p_1\cdot\epsilon+{\cal R}(p_1)\cdot\epsilon}{p_1\cdot k} - 
{\cal D}^\mu(p_1) \frac{\partial}
{\partial p_1^\mu} \right] F_\alpha (\overline{s},\overline{t}) 
\nonumber \\
Y^\alpha\cdot\epsilon & \equiv & 
\left[  
\frac{p_4\cdot\epsilon+{\cal R}(p_4)\cdot\epsilon}{p_4\cdot k} + 
{\cal D}^\mu(p_4) \frac{\partial}
{\partial p_4^\mu} \right] t^\alpha F_\alpha (\overline{s},\overline{t}) 
\nonumber \\
& & - t^\alpha \left[
\frac{p_2\cdot\epsilon+{\cal R}(p_2)\cdot\epsilon}{p_2\cdot k} - 
{\cal D}^\mu(p_2) \frac{\partial}
{\partial p_2^\mu} \right]  F_\alpha (\overline{s},\overline{t})
\label{eq:LOW}
\end{eqnarray}
\item 
\begin{flushleft}
{\sc TsTts} amplitude:
\end{flushleft}
\begin{eqnarray}
X^\alpha\cdot\epsilon & \equiv & 
\left[ \frac{p_3\cdot\epsilon+{\cal R}(p_3)\cdot\epsilon}{p_3\cdot k} - 
\frac{(p_3+p_4)\cdot\epsilon}{(p_3+p_4)\cdot k}
\right] t^\alpha F_\alpha (s_{12},t_{24}) \nonumber \\
& & - t^\alpha
\left[ \frac{p_1\cdot\epsilon+{\cal R}(p_1)\cdot\epsilon}{p_1\cdot k} - 
\frac{(p_1+p_2)\cdot\epsilon}{(p_1+p_2)\cdot k}
\right] F_\alpha (s_{34},t_{24}) \nonumber \\
Y^\alpha\cdot\epsilon & \equiv & 
\left[ \frac{p_4\cdot\epsilon+{\cal R}(p_4)\cdot\epsilon}{p_4\cdot k} - 
\frac{(p_3+p_4)\cdot\epsilon}{(p_3+p_4)\cdot k}
\right] t^\alpha F_\alpha (s_{12},t_{13}) \nonumber \\
& & - t^\alpha
\left[ \frac{p_2\cdot\epsilon+{\cal R}(p_2)\cdot\epsilon}{p_2\cdot k} - 
\frac{(p_1+p_2)\cdot\epsilon}{(p_1+p_2)\cdot k}
\right] F_\alpha (s_{34},t_{13}) 
\label{eq:TSTTS}
\end{eqnarray}
\item 
\begin{flushleft}
{\sc TuTts} amplitude:
\end{flushleft}
\begin{eqnarray}
X^\alpha\cdot\epsilon & \equiv & 
\left[ \frac{p_3\cdot\epsilon+{\cal R}(p_3)\cdot\epsilon}{p_3\cdot k} - 
\frac{(p_2-p_3)\cdot\epsilon}{(p_2-p_3)\cdot k}
\right] t^\alpha F^{'}_\alpha (u_{14},t_{24}) \nonumber \\
& & - t^\alpha
\left[ \frac{p_1\cdot\epsilon+{\cal R}(p_1)\cdot\epsilon}{p_1\cdot k} - 
\frac{(p_1-p_4)\cdot\epsilon}{(p_1-p_4)\cdot k}
\right] F^{'}_\alpha (u_{23},t_{24}) \nonumber \\
Y^\alpha\cdot\epsilon & \equiv & 
\left[ \frac{p_4\cdot\epsilon+{\cal R}(p_4)\cdot\epsilon}{p_4\cdot k} - 
\frac{(p_1-p_4)\cdot\epsilon}{(p_1-p_4)\cdot k}
\right] t^\alpha F^{'}_\alpha (u_{23},t_{13}) \nonumber \\
& & - t^\alpha
\left[ \frac{p_2\cdot\epsilon+{\cal R}(p_2)\cdot\epsilon}{p_2\cdot k} - 
\frac{(p_2-p_3)\cdot\epsilon}{(p_2-p_3)\cdot k}
\right] F^{'}_\alpha (u_{14},t_{13})  
\label{eq:TUTTS}
\end{eqnarray}
\end{itemize}
For the {\sc TuTts} expressions we have defined the functions 
$F^{'}_\alpha(u,t)\equiv F_\alpha(s,t)$ under the $pp$ elastic phase
space constraint $s+t+u=4m^2$.

One can easily see by comparison of Eqs.~(\ref{eq:LOW}), (\ref{eq:TSTTS}) 
and (\ref{eq:TUTTS}) with Eqs.~(\ref{eq:unsym_low}), 
(\ref{eq:unsym_tstts}), and (\ref{eq:unsym_tutts}) respectively
that these results are very similar to those obtained for the spin-0 case. They
each contain the extra factor $R(p)$ which arises from the magnetic moment part
of the electromagnetic coupling. The scalar amplitudes $A$ of the spin-0 cases
are replaced by a sum over terms involving the scalar amplitudes $F_\alpha$,
which are functions of the same variables as $A$, times a momentum
independent matrix factor $t_\alpha$. Because of the Dirac structure, one must
be careful about the ordering of the $t_\alpha$ factors, and of course include
the spinors as in Eq.~(\ref{eq:genspinhalf}). The important point though is
that the dependence on kinematic factors and on the scalar variables appearing
in the amplitudes is essentially the same as in the spin-0 case.

The phase space problem noted for the spinless case in
section~\ref{sectIII} depends only on kinematics, not on the particles' spins,
and so carries over directly to proton-proton bremsstrahlung. 
The physical region of radiative variables pairs such as $(s_{34},t_{24})$
can still lie outside of the measurable region in the $(s,t)$ plane of 
non-radiative phase space.

The amplitudes given above must be antisymmetrized if we 
are to treat identical spin-$\frac{1}{2}$ particle scattering.
The antisymmetrization of the spin-$\frac{1}{2}$ amplitudes is no different
in principle than the symmetrization of spin-$0$ amplitudes given in the 
preceding section and the results are identical: the 
Low-$(\overline{s},\overline{t})$ amplitude can be successfully 
antisymmetrized while still being expressed solely in terms of the measured 
elastic phase shifts; the {\sc TsTts} and {\sc TuTts} amplitudes cannot
be so expressed. We now show these attempts at antisymmetrization explicitly
and demonstrate how the problems arise. 

As has been shown previously for example by Fearing \cite{fearing72b}, one can 
antisymmetrize the Low $(\overline{s}, \overline{t})$ amplitude using 
an analogous procedure to that shown above for $p p$ elastic scattering. 
One would take the amplitude of Eqs.~(\ref{eq:genspinhalf}) 
and~(\ref{eq:LOW}), exchange 
$p_3^\mu \leftrightarrow p_4^\mu$, and apply the Fierz manipulation. The
result is then in the same form as Eq.~(\ref{eq:LOW}), but with 
$F_\alpha(\overline{s},\overline{t})$ replaced by 
$\sum_{\beta=1}^5 {\cal C}_{\beta \alpha} F_\beta(\overline{s}, \overline{u})$.
The final antisymmetrized form is then also identical to Eq.~(\ref{eq:LOW})
but with $F_\alpha(\overline{s},\overline{t})$ replaced by 
$F^{A(Low)}_\alpha(\overline{s},\overline{t})$ where 
\begin{eqnarray}
 F^{A(Low)}_\alpha(\overline{s}, \overline{t}) \equiv 
   F_\alpha(\overline{s}, \overline{t}) - \sum_{\beta=1}^5 
   {\cal C}_{\beta \alpha} F_\beta(\overline{s}, \overline{u}) 
\end{eqnarray}
and $\overline{s}, \overline{t}, \overline{u}$ satisfy the radiative
phase space constraint $\overline{s}+\overline{t}+\overline{u}=4m^2$.
By their identical definitions we see that $F^{A(Low)}_\alpha
\equiv F^A_\alpha$. Operationally, therefore, one only has to take the 
antisymmetrized $F^A_\alpha$ from a phase shift analysis of $p p$
elastic scattering data and insert these functions in the unsymmetrized
Low $(\overline{s}, \overline{t})$ amplitude of Eq.~(\ref{eq:LOW}) 
in order to ensure the correct antisymmetrization of this radiative amplitude.

When we attempt the same procedure for the {\sc TsTts} amplitude a problem 
arises. The calculation goes just as with the $(\overline{s}, \overline{t})$
case except for the definition of the four antisymmetrized $F_\alpha$ 
functions. To obtain the antisymmetrized {\sc TsTts} amplitude we
must replace the $F_\alpha$ of the unsymmetrized amplitude of 
Eq.~(\ref{eq:TSTTS}),
in analogy with Eqs.~(\ref{eq:sym_tstts}) and (\ref{eq:Asym_tstts}),
as follows:
\begin{eqnarray}
F_\alpha(s_{34},t_{24}) & \longrightarrow & 
F^{A(1)}_\alpha(s_{34},t_{24},u_{23})
\equiv F_\alpha(s_{34},t_{24}) - \sum_{\beta=1}^5 {\cal C}_{\beta\alpha}
F_\beta(s_{34},u_{23}) \nonumber \\
F_\alpha(s_{34},t_{13}) & \longrightarrow & 
F^{A(2)}_\alpha(s_{34},t_{13},u_{14})
\equiv F_\alpha(s_{34},t_{13}) - \sum_{\beta=1}^5 {\cal C}_{\beta\alpha}
F_\beta(s_{34},u_{14}) \nonumber \\
F_\alpha(s_{12},t_{24}) & \longrightarrow & 
F^{A(3)}_\alpha(s_{12},t_{24},u_{14})
\equiv F_\alpha(s_{12},t_{24}) - \sum_{\beta=1}^5 {\cal C}_{\beta\alpha}
F_\beta(s_{12},u_{14}) \nonumber \\
F_\alpha(s_{12},t_{13}) & \longrightarrow & 
F^{A(4)}_\alpha(s_{12},t_{13},u_{23})
\equiv F_\alpha(s_{12},t_{13}) - \sum_{\beta=1}^5 {\cal C}_{\beta\alpha}
F_\beta(s_{12},u_{23}).
\end{eqnarray}
The various Lorentz variables in these definitions do not satisfy the
same internal constraints as do the $s, t, u$ of $F^A_\alpha(s, t)$---for 
example, in $F^{A(3)}_\alpha(s_{12}, t_{24}, u_{14})$ we have that
\[s_{12}+t_{24}+u_{14} = 4m^2+2 k\cdot p_3 \neq 4m^2\] in general. 
Thus these $F^{A(1,2,3,4)}_\alpha$
cannot be simply replaced by the $F^A_\alpha(s, t)$ of $p p$ elastic 
scattering as defined in Eq.~(\ref{eq:FAdef}). The functions are not 
identically defined. 
To make such a replacement will result in the radiative amplitude 
having improper symmetry properties. As in the spin-0 case, however, one can
calculate the error introduced by such a replacement. Again the naive estimate
of the error is too pessimistic, due to cancellation, and the actual error is
only of ${\cal O}(k)$.

While attempting to antisymmetrize the {\sc TuTts} amplitude we find 
another difficulty, just as we did in the spin-0 case. 
After replacing $p^\mu_3 \leftrightarrow p^\mu_4$ in the unsymmetrized 
version of {\sc TuTts} and applying the Fierz manipulation we find
\begin{eqnarray}
M^{3 \leftrightarrow 4}_\mu & = & eQ_p \sum_{\alpha=1}^5 \sum_{\beta=1}^5 
{\cal C}_{\beta\alpha} 
\left( F^{'}_\beta (t_{13},u_{23}) \overline{u}_4 \left[ 
\frac{p_{4\mu}+{\cal R}_\mu(p_4)}{p_4\cdot k} - 
\frac{(p_2-p_4)_\mu}{(p_2-p_4)\cdot k}
\right] t_\alpha u_2 
\overline{u}_3 t^\alpha u_1 \right. \nonumber \\
& & \hphantom{eQ_p \sum_{\alpha=1}^5 \sum_{\beta=1}^5 {\cal C}_{\beta\alpha}} 
- F^{'}_\beta (t_{24},u_{23}) \overline{u}_4 t_\alpha u_2 
\overline{u}_3 t^\alpha \left[ \frac{p_{1\mu}+{\cal R}_\mu(p_1)}{p_1\cdot k} -
\frac{(p_1-p_3)_\mu}{(p_1-p_3)\cdot k} \right] u_1 \nonumber \\ 
& & \hphantom{eQ_p \sum_{\alpha=1}^5 \sum_{\beta=1}^5 {\cal C}_{\beta\alpha}} 
+ F^{'}_\beta (t_{24},u_{14}) \overline{u}_4 t_\alpha u_2 
\overline{u}_3 \left[ \frac{p_{3\mu}+{\cal R}_\mu(p_3)}{p_3\cdot k} - 
\frac{(p_1-p_3)_\mu}{(p_1-p_3)\cdot k} \right] t^\alpha u_1 \nonumber \\
& & \hphantom{eQ_p \sum_{\alpha=1}^5 \sum_{\beta=1}^5 {\cal C}_{\beta\alpha}} 
- F^{'}_\beta (t_{13},u_{14}) \overline{u}_4 t_\alpha \left.
\left[ \frac{p_{2\mu}+{\cal R}_\mu(p_2)}{p_2\cdot k} - 
\frac{(p_2-p_4)_\mu}{(p_2-p_4)\cdot k}
\right] u_2 \overline{u}_3 t^\alpha u_1 \right).
\end{eqnarray}
The full antisymmetrized amplitude is then obtained by subtracting this from
the amplitude obtained from Eqs.~(\ref{eq:genspinhalf}) and (\ref{eq:TUTTS}).
The factors $R(p)$ involving the magnetic moments cause no problem. They can be
separated, along with the $p_i/k \cdot p_i$ pieces, from the $F'_\alpha$ which
can then be put in a form analogous to Eq.~(\ref{eq:FAdef}).  The
$(p_i-p_j)^\mu/k \cdot (p_i-p_j)$ terms mix up the momenta however, just as for
the spin-0 case and prevent the radiative amplitudes from being 
put into the original form 
given in Eqs.~(\ref{eq:genspinhalf}) and (\ref{eq:TUTTS}), except with the
antisymmetrized $F'_\alpha$. Thus it is
impossible to express the {\sc TuTts} amplitude purely in terms of the
antisymmetrized $F^A_\alpha$ functions which are the measurable quantities. 
The amplitudes {\sc TsTts} and {\sc TuTts} could 
be correctly antisymmetrized by
computing $M_\mu - M_\mu(p_3 \leftrightarrow p_4)$ directly in terms of the
unsymmetrized functions $F_\alpha(s, t)$. As stated previously however, the 
$F_\alpha(s, t)$ cannot be derived from $p p$ elastic scattering data alone. 
Thus we would lose the direct connection between a process and its 
radiative counterpart which gives the soft photon theorem its utility.
Alternatively, one could modify the original {\sc TuTts} amplitude by changing
the terms added to give gauge invariance, just as was discussed for the spin-0
case.

From Ref.~\cite{liou95} we see that Liou {\it et al.\/} seem to have 
used the antisymmetric $F^A_\alpha(s, t)$ $pp$ elastic functions 
directly in their unsymmetrized {\sc TsTts} and {\sc TuTts} amplitudes. 
Their results cannot have the correct symmetry properties, and again the error
will be of ${\cal O}(k)$ for the {\sc TsTts} amplitude and of 
${\cal O}(k/k)$ for the {\sc TuTts}. 

We shall now illustrate these ideas with some numeric results for the 
$pp$ bremsstrahlung soft
photon spectrum. Low energy proton-proton elastic scattering data is
usually parameterized in terms of a phase shift fit. For our
calculations we have used as input the recent analysis of the Nijmegen group, 
Refs.~\cite{deswart90,deswart93,NN-Online}. The relationship between
these phase shifts and the invariant functions $F_\alpha^A(s,t)$ is
straightforward but rather algebraically involved, and is set down
explicitly in Ref.~\cite{mythesis}. 
In order to investigate the work of Liou, Timmermans and Gibson
\cite{liou95} we have inserted these $F_\alpha^A$ in place of the
unsymmetrized functions $F_\alpha$ in the soft photon amplitudes of
Eqs.~(\ref{eq:LOW}), (\ref{eq:TSTTS})~and~(\ref{eq:TUTTS}). As shown
above this will give rise to the correctly antisymmetrized radiative
amplitude only for the Low-$(\overline{s},\overline{t})$ case of
Eq.~(\ref{eq:LOW}). 

\begin{figure} 
\unitlength1.0cm    
\begin{center}
\begin{picture}(12.0,10) 
\put(-.5,0){\epsfig{file=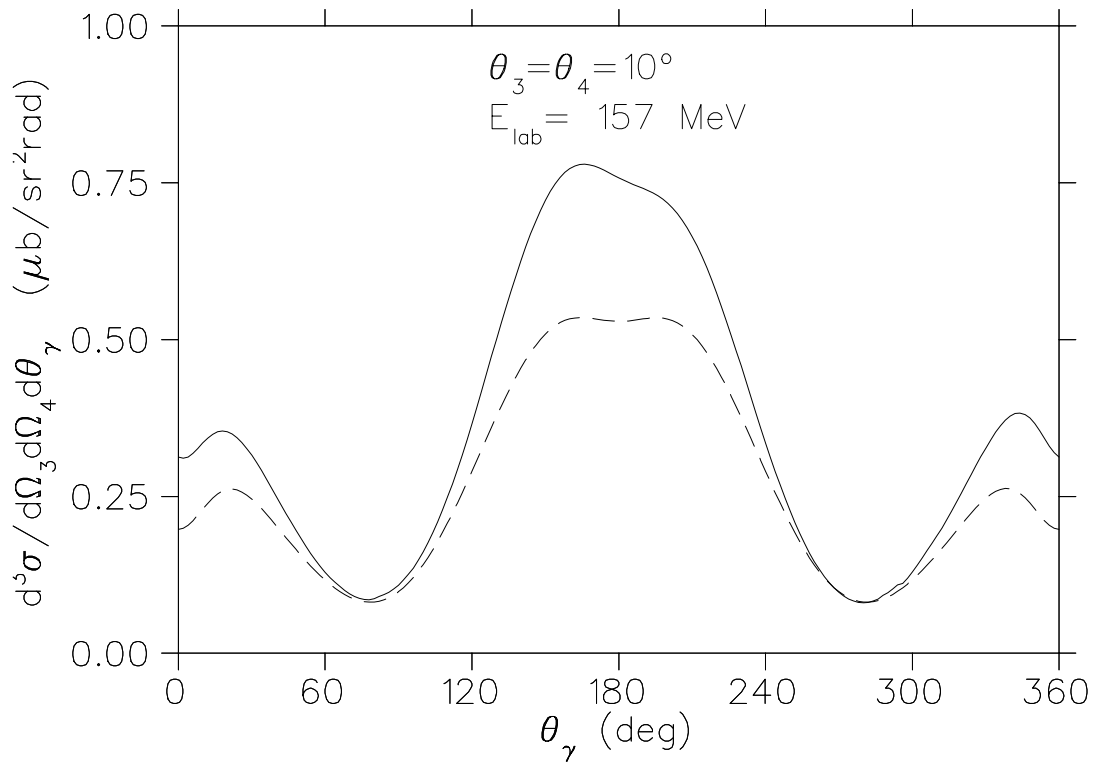,angle=90,height=10cm}}
\end{picture} 
\end{center}
\caption{The solid line shows the {\sc TuTts} soft photon approx\-imation for 
$pp$ bremsstrahlung for the case $T=157\;\mbox{MeV}$, 
$\theta_3=\theta_4=10^\circ$. 
The Low-$(\overline{s},\overline{t})$ soft photon amplitude  
is shown by the dashed line. Both
are computed using the most recent Nijmegen phase shift dataset. 
The spectra should 
exhibit mirror symmetry in $\theta_\gamma$ about the point 
$\theta_\gamma=180^\circ$, since the protons are emitted at equal angles
on either side of the beam direction. The Low-$(\overline{s},\overline{t})$
result is symmetric, but the {\sc TuTts} result is clearly not. This is 
in agreement with our work of section~\protect\ref{sectV}.}
\label{fig:ppbrem}
\end{figure}

In Fig.~\ref{fig:ppbrem} we have shown the differential spectrum
$d^5\sigma / d\Omega_3 d\Omega_4 d\theta_k$ as a function of the
laboratory frame photon angle $\theta_k$ for one of the kinematic
choices studied by the Harvard $pp$ bremsstrahlung experiment of
Ref.~\cite{gottschalk66}---that is, with beam kinetic energy of
157~MeV, with final state protons detected at $10^\circ$ on either
side of the beamline direction, and with all particles coplanar. Shown
are the Low-$(\overline{s},\overline{t})$ and {\sc TuTts} soft photon
calculations using the Nijmegen phase shift dataset as input. From the
symmetry of this experimental setup it is clear that the photon
angular spectrum must be symmetric under reflection about the beamline
axis. This is the case for the Low-$(\overline{s},\overline{t})$
spectrum, while the {\sc TuTts} spectrum does not have this symmetry
property. Since the
authors of Ref.~\cite{liou95} give their results only in the region
$\theta_k = 0^\circ \rightarrow 180^\circ$ this cannot be checked directly 
from their paper. We obtain results shown in Fig.~\ref{fig:ppbrem} 
which agree very well with their results for that 
range of $\theta_k$ but are not mirror symmetric about 
$\theta_k = 180^\circ$, which they should be. This is consistent with an
error in the antisymmetrization of the {\sc TuTts} amplitude. 

Due to the phase space problem described in section~\ref{sectII} we
cannot present a result for the {\sc TsTts} amplitude without
extrapolating the function $F^A_\alpha$ far outside of the physical
region of non-radiative phase space. In terms of the phase shifts,
which are parameterized by center-of-mass frame momentum and
scattering angle $\theta_{CM}$, this would correspond to evaluating
the non-radiative amplitude for $\cos\theta_{CM}<-1$. The authors of
Ref.~\cite{liou95} do present results for the {\sc TsTts}
amplitude. These spectra differ by large factors from other
soft photon calculations, from potential model calculations, and from
experimental data. This difference appears in spite of the well known
result that the leading two orders of the photon spectrum
are uniquely predicted. The authors of Ref.~\cite{liou95} suggest that
this large discrepancy is evidence that the {\sc TuTts} amplitude
should be preferred over the {\sc TsTts} amplitude for calculations of
identical particle scattering. An alternative explanation might be that this
large discrepancy is simply a reflection of the fact that the elastic
amplitudes, which are required far outside the physical region, were
extrapolated in some unphysical way.

\section{Conclusions}
\label{sectVI}

We have seen that problems arise in the practical application of
certain soft photon amplitudes to two-body bremsstrahlung processes,
in particular to $pp$~bremsstrahlung. Since the variables
$\overline{s}$, $\overline{t}$ and~$\overline{u}$ of radiative phase
space satisfy the same constraints as the Mandelstam variables $s$,
$t$ and~$u$ of the corresponding non-radiative process we find that
the usual Low formulation of the approximation, which expresses all elastic
amplitudes at a single point in terms of these variables, is unaffected by the
phase space problem and the antisymmetrization problem. In contrast,
neither the ``Two-u-Two-t-special'' ({\sc TuTts}) nor the
``Two-s-Two-t-special'' ({\sc TsTts}) amplitudes suggested in
Ref.~\cite{liou95} can be antisymmetrized while being written in terms
of the measurable $pp$~elastic amplitude. Additionally the {\sc TsTts}
amplitude was found to be incalculable unless one makes a
model-dependent extrapolation of the $pp$~elastic amplitude outside of
its physical, on-shell region. We conclude that the {\sc TuTts} and
{\sc TsTts} soft photon amplitudes, and those other alternative forms for
the soft photon approximation which rely on Taylor expansions about
radiative variable pairs other than $(\overline{s},\overline{t})$ and share the
same problems, are
not suitable for the soft photon analysis of proton-proton
bremsstrahlung. 

\section{Acknowledgments}

This work was supported in part by a grant from the Natural Sciences and
Engineering Research Council of Canada.

\newpage

\end{document}